\documentstyle[12pt]{article}

\def\Z{{\bf Z}}

\def\C{{\bf C}}
\def\a{\alpha}
\def\aa{\langle a \rangle}
\def\b{\beta}
\def\m{\mu}
\def\n{\nu}
\def\tm{\tilde{m}}
\def\th{\theta}
\def\tr{{\rm Tr}}
\def\pf{{\rm Pf}}
\def\g{\gamma}
\def\G{\Gamma}
\def\O{\Omega}

\def\P{\Phi}
\def\teP{\varphi}

\def\l{\lambda}
\def\L{\Lambda}
\def\D{\Delta}

\def\pa{\partial}
\def\p{\phi}

\def\e{\epsilon}
\def\z{\zeta}
\def\t{\tau}
\def\implies{\Rightarrow}
\def\d{\delta}
\def\bi{\bar{i}}
\def\bQ{\bar{Q}}
\def\diag{{\rm diag}}

\begin{document}

\newcommand{\inv}[1]{{#1}^{-1}} %inverse

\renewcommand{\theequation}{\thesection.\arabic{equation}}
\newcommand{\beq}{\begin{equation}}
\newcommand{\eeq}[1]{\label{#1}\end{equation}}
\newcommand{\ber}{\begin{eqnarray}}
\newcommand{\eer}[1]{\label{#1}\end{eqnarray}}
%\begin{titlepage}
\begin{center}
  August, 1995
                                \hfill    RI-7-95\\
                                \hfill    hep-th/9509130\\

\vskip .4in

{\large \bf Effective Potentials and Vacuum Structure
in $N=1$ Supersymmetric Gauge Theories}
\vskip .4in

{\bf Shmuel Elitzur}, \footnotemark\
\footnotetext{e-mail address: elitzur@vms.huji.ac.il}
{\bf Avraham Forge}, \footnotemark\
\footnotetext{e-mail address: forge@vms.huji.ac.il}
{\bf Amit Giveon}, \footnotemark\
\footnotetext{e-mail address: giveon@vms.huji.ac.il}

\vskip .1in

{\em Racah Institute of Physics, The Hebrew University\\
  Jerusalem, 91904, Israel}

\vskip .1in

{\bf Eliezer Rabinovici} \footnotemark\
\footnotetext{Permanent address: Racah Institute of Physics, The Hebrew
University, Jerusalem, 91904, Israel; e-mail address: eliezer@vms.huji.ac.il}

\vskip .1in

{\em Department of Physics, Rutgers University\\ Piscataway, NJ 08855-0849,
USA }

\end{center}
\vskip .3in
\begin{center} {\bf ABSTRACT } \end{center}
\begin{quotation}\noindent

We derive the exact effective superpotential in $4d$, $N=1$ supersymmetric
$SU(2)$ gauge theories with $N_A$ triplets and $2N_f$ doublets of matter
superfields.
We find the quantum vacua of these theories; the equations of motion
(for $N_A=1$) can be reorganized into the singularity conditions of an
elliptic curve.
{}From the phase transition points to the Coulomb branch,
we find the exact Abelian gauge couplings, $\tau$, for arbitrary
bare masses and Yukawa couplings. We thus {\em derive} the result that
$\tau$ is a section of an $SL(2,\Z)$ bundle over the moduli space and over
the parameters space of bare masses and Yukawa couplings.
For $N_c>2$, we derive the exact effective superpotential in branches of
supersymmetric $SU(N_c)$ gauge theories with one supermultiplet in the
adjoint representation ($N_A=1$) and  zero or one flavor ($N_f=0,1$).
We find the quantum vacua of these theories; the equations of motion
can be reorganized into the singularity conditions of a genus $N_c-1$
hyperelliptic curve.
Finally, we present the effective superpotential in the $N_A$, $N_f<N_c$
cases.

\end{quotation}
\vfill
\eject
\def\baselinestretch{1.2}
\baselineskip 16 pt

\noindent
\section{Introduction and Discussion}
\setcounter{equation}{0}
Recently, many new exact results were derived in four dimensional
supersymmetric gauge theories (for a review, see ref. \cite{S} and
references therein).
In particular, in ref. \cite{efgr}, we reported the results of applying
the methods of refs. \cite{S,ILS,I,IS1} to the general case of an infra-red
nontrivial $N=1$ supersymmetric $SU(2)$ gauge theory with $N_A$ triplets
of matter superfields and $N_f$, $N=2$ flavors ({\em i.e.}, $2N_f$
doublets). In \cite{efgr}, we presented the exact effective superpotential
in these models, and the effective Abelian gauge couplings for arbitrary
bare masses and Yukawa couplings.

In this paper, we present the detailed derivation of the results of ref.
\cite{efgr}. Moreover, we extend the results to the ``$SU(N_c)$ vacua''
branches of supersymmetric $SU(N_c)$ gauge theories with $N_A$ matter
superfields in the adjoint representation, $N_f$ supermultiplets in the
fundamental and $N_f$ supermultiplets in the anti-fundamental
representations ($N_f<N_c$).

To derive the nonperturbative superpotential of a particular supersymmetric
gauge theory, one may attempt to obtain a unique result by using
holomorphy, symmetries and
limiting considerations \cite{S}. An equivalent, but more efficient way,
in some cases, is to derive the exact superpotential by
applying similar considerations in an ``integrating in'' procedure
\cite{ILS,I}.
Under certain conditions, one may, unconventionally, derive the effective
superpotential for modes which are of finite mass, given the effective
action in which these modes have been considered to have infinite mass.
In this work, we apply the integrating in technique when it is valid; the
various consistency checks to which the result is subjected strengthen the
reliability of the method.

We begin, in section 2, with a review of the integrating in procedure. We
discuss the limiting considerations which determine when such a procedure
allows to derive the nonperturbative superpotential. Then, in section 3,
we review the integrating in of $2N_f$ doublets to pure $N=1$
supersymmetric $SU(2)$ gauge theory.

In section 4, we integrate in $N_A$ triplets to an $N=1$ supersymmetric
$SU(2)$ gauge theory with $2N_f$ doublets. We find a universal
representation of the nonperturbative superpotential for all (one-loop)
infra-red nontrivial theories, namely, with $N_A=3$, $N_f=0$, or
$N_A=2$, $N_f=0,1,2$, or $N_A=1$, $N_f=0,1,2,3,4$. We then review the
physics of cases without doublets ($N_f=0$).

In sections 5,6,7,8, we consider in detail the $SU(2)$ models with one
triplet matter superfield ($N_A=1$), and with $N_f=1,2,3,4$, respectively.
In all cases, we find the quantum vacua in the Higgs/confinement \cite{BR}
branches
(the ``$SU(2)$ vacua''). We reorganize some of the equations, derived by
variations of the superpotential, into the singular conditions of an elliptic
curve. This elliptic curve has $2+N_f$ singularities, corresponding to the
$2+N_f$ branches of $SU(2)$ vacua; the values of the quantum field,
corresponding classically to the $SU(2)$ quadratic Casimir, are fixed.
The rest of the equations of motion
determine the values of other quantum fields as functions of the bare masses
and Yukawa couplings in each branch. In the massless case,
there is a $Z_{4-N_f}$ global symmetry~\footnote{
$Z_1$ means no symmetry and $Z_0$ means $U(1)$.}
acting on the moduli space of the $N_A=1$, $N_f>0$ theory; this symmetry
can be read directly from the quantum superpotential.

Moreover, at the phase transition points to the Coulomb branch, the results
in sections 5,6,7,8 provide a direct derivation of the elliptic curves,
defining the effective Abelian gauge coupling, $\tau$,
as a function of the bare masses and Yukawa couplings.
Therefore, we {\em derive} the result that $\tau$ is a modular parameter of
a torus, namely, a section of an $SL(2,\Z)$ bundle over the moduli space
and over the parameters space of bare masses and Yukawa couplings.

These results pass various  consistency
checks (like integrating out of any degree of freedom). In particular,
on the subspace of parameters where the theory has an
enhanced $N=2$ supersymmetry, we reobtain the results of Seiberg and Witten
\cite{SW1,SW2}. On the way, we identify a physical meaning of the
complex parameter $x$ appearing in the elliptic curve equation: $y^2=p(x)$.

At each of the $2+N_f$ singular points, in the moduli space of the Coulomb
branch, a dyon becomes massless. For special values of the bare masses and
Yukawa couplings, some of the $2+N_f$ vacua degenerate. In some cases, it
may lead to points where mutually non-local degrees of freedom are
massless~\footnote{We thank R. Plesser and N. Seiberg for discussions on
this point.}, similar to the situation in pure $N=2$ supersymmetric $SU(3)$
gauge theories, considered in \cite{ad}. For example, when all masses and
Yukawa couplings approach zero, all the $2+N_f$ singularities collapse to
the origin. Such a point might be interpreted as a non-Abelian Coulomb
phase \cite{S2}.

In sections 9,10, we consider the $SU(2)$ models with two triplet matter
superfields ($N_A=2$), and with $N_f=1,2$. In section 9, we find the
quantum vacua of the model with one flavor ($N_f=1$). Again, at the phase
transition points to the Coulomb branch, the equations derived by
variation of the superpotential can be reorganized into the singular
conditions of an elliptic curve -- the one defining the effective Abelian
gauge coupling in the Coulomb branch. Unlike the $N_A=1$ cases, away from
the phase transition points, the equations determining the vacua
fail to describe, in general, an elliptic curve.  As before, on the
way, we identify a physical meaning of the complex parameter $x$, which
at the phase transition points becomes the one appearing in the elliptic
curve equation: $y^2=p(x)$. For special values of the bare parameters, we
consider a point in the moduli space that might be interpreted as a
non-abelian Coulomb phase \cite{S2} and, as another consistency check,
we show that the reduction from $N_A=2$ to $N_A=1$ is simple.

In section 10, we argue that the supersymmetric $SU(2)$ gauge theory with
$N_A=N_f=2$ is infra-red free. This result is consistent with the fact that
(unlike the other one-loop conformal cases: $N_A=1$, $N_f=4$, or $N_A=3$,
$N_f=0$) we are not able to match the coupling constant of this theory to the
one of the model  with $N_A=0$, $N_f=2$, in a way that respects the global
symmetries of the theory. The general discussion in section 10 follows
ref. \cite{LS}. Analyzing the gauge-coupling beta-function and the Yukawa
couplings beta-functions, we see that in the other one-loop infra-red
conformal theories, indeed, the beta-function equations are expected to
have a fixed line of solutions.

In sections 11,12,13, we turn to supersymmetric $SU(N_c)$ gauge theories
with more than two colors: $N_c>2$. In section 11, we consider $SU(N_c)$
with one matter superfield in the adjoint representation ($N_A=1$)
and no flavors ($N_f=0$). In this case,
symmetries and limiting considerations are not strong enough to allow a
determined integrating in procedure from the pure supersymmetric
$SU(N_c)$ gauge theory (although they are enough to show that the
effective superpotential vanishes). However, imposing also the physical
condition to have a discrete set of ``$SU(N_c)$ vacua,'' allows one to obtain
the result: the nonperturbative
superpotential vanishes and there are constraints which fix
the quantum field, corresponding classically to the $SU(N_c)$
quadratic Casimir, at $N_c$ values,
and fix the value of quantum fields, corresponding classically
to higher Casimirs, to zero. These are exactly the $N_c$ $SU(N_c)$ vacua of
the theory; there is a $Z_{N_c}$ global symmetry, acting in the moduli space,
which relates them.

In section 12, we present the nonperturbative superpotential in
supersymmetric $SU(N_c)$ gauge theory with $N_f$ flavors and $N_A=0$, found
in refs. \cite{study,S1}. Then, in section 13, we add to the
$N_f=1$ theory a matter superfield in the adjoint representation.
As before, symmetries and limiting considerations are not strong enough to
allow a determined integrating in procedure from the supersymmetric
$SU(N_c)$ gauge theory with $N_f=1$ to the theory with $N_A=N_f=1$.
However, imposing also the physical
condition to have a discrete set of $SU(N_c)$ vacua, allows one to obtain
the result: the nonperturbative superpotential does not depend on the
quantum fields, corresponding classically to $SU(N_c)$ Casimirs, except for
the quadratic Casimir; the quantum fields corresponding to higher Casimirs
are constrained.

We find the quantum vacua in the fully Higgsed/confined branches
(the ``$SU(N_c)$ vacua'') in the presence of a tree-level superpotential
containing only mass terms and a Yukawa coupling term. We reorganize some of
the equations, derived by
variations of the superpotential, into the singular conditions of a genus
$N_c-1$ hyperelliptic curve. This hyperelliptic curve has $2N_c-1$
singularities, corresponding to the $N_c+(N_c-1)N_f=2N_c-1$ $SU(N_c)$ vacua
of the $N_A=N_f=1$ theory.
In the massless case, there is a $Z_{2N_c-1}$ global symmetry
acting on the moduli space of this theory; this symmetry
can be read directly from the quantum superpotential.

In section 14, we revisit the supersymmetric $SU(2)$ gauge theory with
$N_A=N_f=1$, in a way similar to the manipulation for $N_c>2$.
We find that the equations of motion, determining the $SU(2)$ vacua,
are the singularity conditions of an elliptic curve, related to the
previous one by an $SL(2,\C)$ transformation.

In section 15, we present the effective superpotential in
supersymmetric $SU(N_c)$ gauge theories, $N_c>2$, with $N_A$ matter
superfields in the adjoint representation and $N_f<N_c$ flavors. For
$N_A=1$, we find that there are $N_c+N_f(N_c-1)-{1\over 2}N_f(N_f-1)$
(branches of) $SU(N_c)$ vacua in the presence of a tree-level superpotential
containing only mass terms and Yukawa coupling terms. In the massless case,
there is a $Z_{2N_c-N_f}$ global symmetry
acting on the moduli space of the $N_A=1$, $N_f>0$ theory; this symmetry
can be read directly from the quantum superpotential.
For $N_A=2$, $N_f\neq 0$, the superpotential in section 15 is conjectured.

Finally, in the Appendix, we show in detail the considerations leading
to the conclusion that the integrating in procedure is valid in
examples discussed in this work.

\noindent
\section{Integrating in}
\setcounter{equation}{0}
For completeness, we start by repeating the general discussion of refs.
\cite{ILS,I}. Let the nonperturbative superpotential of an $N=1$
supersymmetric gauge theory, which we call the ``down'' theory, be
\beq
W_d\equiv W_{down}(X_I,\L_d) .
\eeq{Wd}
$W_d$ depends on the gauge singlets $X_I$, which are constructed out of the
down theory matter superfields, $D_i$, and on the gauge coupling constant,
expressed as the dynamically generated scale of the down theory, $\L_d$;
we may add to $W_d$ the tree-level superpotential: $m^I X_I$.

Suppose we also know the nonperturbative superpotential of the $N=1$ gauge
theory with a matter superfield $U$ in addition to $D_i$. We call this
theory the ``up'' theory, and its nonperturbative superpotential is
\beq
W_u\equiv W_{up}(X_I,M,Z_A,\L),
\eeq{Wu}
where $M\sim U^2$ and $Z_A$ are relevant gauge singlets constructed out
of $U$ and $D_i$. In eq. (\ref{Wu}), $\L\equiv \L_u$ is the dynamically
generated scale of the up theory. One may add to $W_u$ the tree-level
superpotential
\beq
W_{tree}=\tm M + \l^A Z_A .
\eeq{Wt}
Here $\tm$ is the mass of the superfield $U$, and $\l^A$ are some couplings
of $U$ and $D_i$.

By integrating out $U$ at finite mass, $\tm$, one gets, using the notation
of ref. \cite{I},
\beq
[W_u+\tm M+\l Z]_{\langle M\rangle, \langle Z\rangle}=W_d+W_{i},
\eeq{io}
where
\beq
W_i\equiv W_{intermediate}(X,\tm,\l).
\eeq{Wi}
We may now split the ``intermediate'' superpotential into two pieces:
\beq
W_i=W_{tree,d}+W_{\D},
\eeq{WD}
where the tree-level down superpotential, $W_{tree,d}$, is
\beq
W_{tree,d}\equiv W_{tree}|_{\langle U\rangle}.
\eeq{Wtd}

So far we have described the obvious ``integrating out'' of $U$
from the up theory to the down theory. Now, suppose we start with the down
theory, and add to it the superfield $U$. One may attempt to obtain a
unique result for $W_u$ by symmetry and limiting considerations. A
more efficient way, in some cases, is to derive $W_u$ by
applying similar considerations in an ``integrating in'' procedure.

The integrating in of $U$
from the down theory to the up theory is possible if we know
$W_{\D}$.\footnote{A list of warnings as to limitations of the procedure
appears in \cite{IS3}.} The superpotential of the up theory
is derived from the superpotential of the down theory by the Legendre
transform of eq. (\ref{io}):
\beq
W_u=[W_d+W_{\D}+W_{tree,d}-W_{tree}]_{\langle\tm\rangle,\langle\l\rangle}.
\eeq{ii}

How can one find $W_{\D}$? It might happen that holomorphy, symmetries, and
various limits are strong enough to impose $W_{\D}=0$. The limits that
$W_{\D}$ should obey are
\beq
W_{\D}(\L,\tm\to \infty)\to 0, \qquad W_{\D}(\L\to 0,\tm)\to 0 .
\eeq{limits}
When $\L\to 0$, the theory becomes classical and the superpotential
collapses to $W_{tree,d}$ (in some cases there is also a classical
constraint). When
$\tm\to\infty$, the additional degrees of freedom $U$ become much heavier
than the scale $\L$ of the up theory and hence are expected to decouple
from the down theory, except for their influence on the renormalization of
the coupling. This will make $W_d$ depend on the down scale $\L_d$.
If $W_{\D}$ is indeed zero, we can integrate in the superfield $U$ and
derive the nonperturbative superpotential of the up theory.

We should remark that the $W_u$, derived by integrating in, is the
superpotential on particular branches in the moduli space of $N=1$ vacua --
those branches which contain the heavy $U$ region. Moreover, $W_u$
is expected to be singular at points in the moduli space where extra
degrees of freedom -- not included in the procedure -- become massless.
We shall return to these points later.

\noindent
\section{Integrating in $2N_f$ doublets to pure $N=1$, $SU(2)$  gauge
theory}
\setcounter{equation}{0}
The nonperturbative superpotential of
$N=1$ supersymmetric $SU(2)$ gauge theory
with $N_f$ flavors ($2N_f$ doublets) can be constructed just by the use of
holomorphy and symmetries \cite{study,S1}. Yet, following refs. \cite{ILS,I},
we shall also derive $W_{N_f}$ by integrating in $N_f$ flavors to the pure
supersymmetric  $SU(2)$ gauge theory.

The classical low-energy effective superpotential of the down theory vanishes,
and the nonperturbative effective superpotential (due to gluino condensation)
is
\beq
W_d({\rm pure}\,\, N=1, SU(2))=\pm 2\L_d^3,
\eeq{Wsu2}
where $\L_d$ is the dynamically generated scale of the down theory.
We now want to integrate in $2N_f$ supermultiplets in the fundamental
representation, $Q_i^a$, $i=1,...,2N_f$, and $a$ is a fundamental
representation index. One-loop asymptotic freedom or
conformal invariance implies
\beq
b_1=6-N_f\geq 0,
\eeq{b1}
where $-b_1$ is the one-loop coefficient of the gauge coupling beta-function.
We consider these models in the presence of masses $m$.
The relevant gauge singlets, $X_{ij}=-X_{ji}$, are quadratic in the $N=1$
superfield doublets, $Q^a$:
\beq
X_{ij}=\e_{ab}Q_i^a Q_j^b, \qquad  a,b=1,2, \qquad i,j=1,...,2N_f,
\eeq{Xij}
and, therefore,
\beq
W_{tree}={1\over 2}\tr_{2N_f}m X \Rightarrow W_{tree,d}\equiv
{1\over 2}\tr_{2N_f} m X|_{\langle Q \rangle}=0.
\eeq{tmX}

By using the global symmetries, $SU(2N_f)\times U(1)_Q\times U(1)_R$,
one finds that $W_{\D}=0$ and, therefore,
\beq
W_u(X)=[W_d-W_{tree}]_{\langle m\rangle}=
[\pm 2(\pf m)^{{1\over 2}} \L^{{6-N_f\over 2}}
-{1\over 2}\tr_{2N_f} m X]_{\langle m\rangle}.
\eeq{WuX}
Here we used the coupling constant matching
(consistent with global symmetries)
\beq
\L_d^{b_{1,d}}=(\pf m)\L^{b_{1,u}},
\eeq{LdL}
where $b_{1,d}=6$ is minus the one-loop coefficient of the gauge coupling
beta-function of the  down theory, and $b_{1,u}=6-N_f$ is minus the
one-loop coefficient of the gauge coupling beta-function of the up theory.
This matching implies $W_d(m,\L)=\pm 2(\pf m)^{{1\over 2}}\L^{{6-N_f\over
2}}$, which we inserted in (\ref{WuX}).
Finally, one finds
\beq
W_{N_f}(X)=(2-N_f)\L^{{6-N_f\over 2-N_f}} (\pf X)^{{1\over N_f-2}}
+{1\over 2}\tr_{2N_f} mX ,
\eeq{WNf}
where an additional tree-level superpotential has been added to $W_u$.
For $N_f=1$, the massless superpotential reads:
$W=\L^5/X$. For $N_f=2$ ($b_1=4$ in eq. (\ref{b1})), $W=0$, and by
the integrating in procedure we also get the constraint: $\pf X=\L^4$. For
$N_f>2$, $W(m=0)$ is proportional to some positive power of the classical
constraint: $\pf X=0$. Small values of $\L$ imply a semi-classical limit for
which the classical constraint is imposed; however, quantum corrections
remove the constraint. At the $\langle X\rangle=0$ vacuum one expects to find
extra massless interacting scalars (by 't Hooft matching conditions
\cite{S1} and, for $N_f>3$, by electric-magnetic duality \cite{S2}) and,
therefore, we make
use of the $N_f>2$ superpotential only in the presence of a mass matrix
$m$ ($\det m\neq 0$), which fixes the vacua at $\langle X\rangle\neq 0$.

\noindent
\section{Integrating in $N_A$ triplets to $N=1$, $SU(2)$  gauge
theory with $2N_f$ doublets}
\setcounter{equation}{0}
We now want to derive the nonperturbative superpotential, $W_{N_f,N_A}$,
of $N=1$ supersymmetric $SU(2)$ gauge
theory with $N_A$ triplets and $N_f$ flavors, by integrating in $N_A$
triplets, $\P_{\a}^{ab}$, $\a=1,...,N_A$,
to the supersymmetric gauge theory with $2N_f$ doublets (\ref{WNf}).
Here $a,b$ are fundamental representation indices,
and $\P^{ab}=\P^{ba}$. As before, we treat the cases with one-loop
asymptotic freedom or conformal invariance, for which
\beq
b_1=6-N_f-2N_A\geq 0,
\eeq{bfA}
where $-b_1$ is the one-loop coefficient of the gauge coupling
beta-function.

The relevant gauge singlets we should add to $X_{ij}$ in eq. (\ref{Xij})
are $M_{\a\b}=M_{\b\a}$ and $Z_{ij\,\a}=Z_{ji\,\a}$:
\ber
M_{\a\b}&=&\e_{aa'}\e_{bb'}\P_{\a}^{ab}\P_{\b}^{a'b'},  \qquad  a,b=1,2,
\qquad \a ,\b=1,...,N_A,
\nonumber\\
Z_{ij\,\a}&=&\e_{aa'}\e_{bb'}Q_i^a\P_{\a}^{a'b'}Q_j^b \qquad
i,j=1,...,2N_f.
\eer{XMZ}

After some algebra one can show that
\ber
W_{tree,d}\,\,\equiv\,\, W_{tree}|_{\langle\P_{\a}\rangle}
&=&[\tr_{N_A}\tm M+{1\over \sqrt{2}}
\tr_{2N_f}\l^{\a}Z_{\a}]_{\langle\P_{\a}\rangle}\nonumber\\
&=&{1\over 8}\tr_{2N_f}(\tm^{-1})_{\a\b} \l^{\a} X \l^{\b} X .
\eer{Wtdfa}
Moreover, by using the global symmetries and various limits one can show that
$W_{\D}=0$; this is done in the Appendix. Therefore,
\ber
W_u(M,X,Z)&=&[W_d+W_{tree,d}-W_{tree}]_{\langle \tm\rangle,\langle\l\rangle}
\nonumber\\&=& \Big[
{2-N_f\over \L^{6-N_f-2N_A\over N_f-2}}(\det\tm)^{-2\over N_f-2}(\pf
X)^{1\over N_f-2}
\nonumber\\&+&{1\over 8}\tr_{2N_f}(\tm^{-1})_{\a\b} \l^{\a} X \l^{\b} X
\nonumber\\&-&\tr_{N_A}\tm M-{1\over \sqrt{2}}\tr_{2N_f}
\l^{\a} Z_{\a}\Big]_{\langle \tm\rangle, \langle\l\rangle}
\eer{WMXZ}
Here we used $W_d\equiv W_{N_f}$ of eq. (\ref{WNf}), where we inserted the
matching  (consistent with global symmetries):
\beq
\L_d^{b_{1,d}}=[\det (\tm/2)]^2\L^{b_{1,u}}.
\eeq{LdL2}
In (\ref{LdL2})  $b_{1,d}=6-N_f$ is minus the one-loop coefficient of the
gauge coupling
beta-function of the  down theory, and $b_{1,u}=6-N_f-2N_A$ is minus the
one-loop coefficient of the gauge coupling beta-function of the up theory.
Finally, we obtain the superpotential\footnote{
When $b_1=4$, the nonperturbative superpotential vanishes and
one also obtains constraints.
In the conformal case, when $b_1=0$, ``$\L^{-b_1}$'' in (\ref{WfA}) should be
replaced by a function of $\tau_0={\th_0\over\pi}+{8\pi i\over g_0^2}$
(the non-Abelian gauge coupling) and $\det \l$; this will be discussed in
section 8.}
\ber
W_{N_f,N_A}(M,X,Z) &=& (b_1-4)\Big\{\L^{-b_1} \pf X\Big[
{\rm det}_{N_A}(\Gamma_{\a\b})\Big]^2
\Big\}^{1/(4-b_1)}\nonumber\\
&+&\tr_{N_A} \tm M +{1\over 2}\tr_{2N_f} mX
+{1\over\sqrt{2}}\tr_{2N_f} \l^{\a} Z_{\a} ,
\eer{WfA}
where
\beq
\Gamma_{\a\b}(M,X,Z)=M_{\a\b}+\tr_{2N_f}(Z_{\a}X^{-1}Z_{\b}X^{-1}).
\eeq{G}
Recall that in eq. (\ref{WfA}),
$\L$ is the dynamically generated scale, $b_1$ is given in eq.
(\ref{bfA}), and $\tm_{\a\b}$, $m_{ij}$
and $\l^{\a}_{ij}$ are the bare masses and Yukawa couplings, respectively
($\tm_{\a\b}=\tm_{\b\a}$, $m_{ij}=-m_{ji}$, $\l^{\a}_{ij}=\l^{\a}_{ji}$).

{}From eqs. (\ref{Xij}), (\ref{XMZ}) it is clear that the determinant in
$W_{N_f,N_A}$ vanishes classically.
Quantum mechanically, the constraint is removed; by taking the $\L\to 0$
limit in eq. (\ref{WfA}), one recovers the classical constraint
${\rm det}_{N_3}(\Gamma_{\a\b})=0$ (if $b_1<4$, $N_f\neq 0$).

Models without triplets ($N_A=0$) were discussed in section 3.
Models without doublets ($N_f=0$) were studied in \cite{SW1,IS1,IS2}.
In these cases
\beq
W_{0,N_A}(M)=
2(1-N_A)\L^{{N_A-3\over N_A-1}}(\det M)^{{1\over N_A-1}}+\tr_{N_A}\tm M.
\eeq{WM}
The massless $N_A=1$ case is a pure $SU(2)$,
$N=2$ supersymmetric Yang-Mills theory.
This model was considered in detail in ref. \cite{SW1}. In this case, $W=0$
(compatible with eq. (\ref{WfA})). As in the other $b_1=4$ case, discussed
in section 3,
by the integrating in procedure one also gets a constraint in this
case: $M=\pm \L^2$. This result can be understood because the starting
point of the integrating in procedure is a pure $N=1$ supersymmetric
Yang-Mills theory. Therefore, it
leads us to the points at the verge of confinement in the moduli space.
These are the two singular points in the
$M$ moduli space of the theory; they are due to massless monopoles or
dyons. Such excitations are not constructed out of the elementary
degrees of freedom and, therefore, there is no trace for them in $W$. (This
situation is different if $N_f\neq 0$; in this case, monopoles are
different manifestations of the elementary degrees of freedom.)

The $N_f=0$, $N_A=2$ case is discussed in refs. \cite{IS1,IS2}. In this
case, the superpotential in eq. (\ref{WfA}) is the one presented in
\cite{IS1,IS2} on the confinement and the oblique confinement branches (it so
happens that for $\tm$ such that $\det \tm=0$ eq. (\ref{WM}) also describes
the Coulomb phase). As in the $N_A=1$ case, this is because the starting
point of the integrating in procedure is a pure $N=1$ supersymmetric
Yang-Mills theory and, therefore, it leads us to the confining branches in
the moduli space. The moduli space may also contain a non-Abelian Coulomb
phase at the point $\langle M\rangle =0$ \cite{IS1,IS3}.

For $N_A=3$ there is an
additional Yukawa coupling that we did not consider in (\ref{XMZ}):
the one which couples the three (antisymmetric) triplets.
Therefore, we should also integrate in the
additional gauge singlet $\P\P\P\equiv \det\P$. The superpotential in
eq. (\ref{WM}) remains valid also in the presence of
$W_{tree}=\l\det\P$ because $\det\P=(\det M)^{1/2}$. For this term,
the Yukawa coupling, $\l$, replaces ``$\L^0$'' in eq. (\ref{WM}).
This result coincides with the one derived in \cite{IS2}.
In the massless case, this
theory flows to an $N=4$ supersymmetric Yang-Mills fixed point.

In sections 5-10, we consider the supersymmetric $SU(2)$ models with
$N_A\neq 0$ and $N_f\neq 0$. All the symmetries and quantum numbers of the
various parameters, in particular, such as used in \cite{SW1,SW2}, are
already embodied in the superpotential $W_{N_f,N_A}$ of eq. (\ref{WfA}).

By construction, integrating out a triplet from $W_{N_f,N_A}$ (\ref{WfA})
gives the superpotential of the down theory: $W_{N_f}$ of eq. (\ref{WNf}).
Moreover, integrating out a flavor from $W_{N_f,N_A}$ gives the
superpotential $W_{N_f-1,N_A}$ and an intermediate superpotential, $W_i$,
which vanishes in the infinite mass limit of the doublets integrated out:
$W_i(m_{N_f-1\, N_f}\to\infty)\to 0$.

We should remark that the singularities at $X=0$, and the branch
cuts in $\pf X$ and $\G$,
signal the appearance of extra massless degrees of freedom at these points
(the branch cuts in $\L$ are due to non-Abelian effects).
Those are expected, physically, due to some duality, similar to
the electric-magnetic duality of refs. \cite{S2,K}.
The $SU(2)$, $N_A=1$, $N_f$ models fall into a lacuna in
the analysis in ref. \cite{K} of the dual models to $SU(N_c)$ systems with
matter in the adjoint and fundamental representations. The results obtained
here might shed some light on this gap.

Finally, to complete the survey of $SU(2)$ models obeying $b_1\geq 0$,
let us note that one can also have an infra-red non-trivial theory
with a single matter superfield in the $I=3/2$ representation.
The $N_{3/2}=1$, $N_f=0$ theory was shown to have $W=0$
\cite{ISS}. Adding $N_f=1$ matter results with $b_1=0$ in eq. (\ref{bfA}).
The two-loop beta function renders the theory infra-red free. As no Yukawa
coupling is possible, this model is indeed infra-red free.

\noindent
\section{$SU(2)$ with $N_A=N_f=1$ ($b_1=3$)}
\setcounter{equation}{0}
Before turning to the substance of this section, we should remark that we
will use some notational and algebraic complications which are not
necessary in the study of the $N=1$, $SU(2)$ case with $N_A=N_f=1$
(we shall return to a simpler manipulation of this case after discussing
$SU(N_c)$ with one adjoint and one flavor in section 14).
We do the manipulation here in a way similar to what we shall do in the
more complicated cases of $SU(2)$ with one adjoint and several flavors.

A supersymmetric $SU(2)$ gauge theory with one triplet and one flavor has a
superpotential (\ref{WfA}):
\beq
W_{1,1}=-\L^{-3}(\pf X)\G^2+\tm M +{1\over 2}\tr\, mX +
{1\over \sqrt{2}} \tr\, \l Z.
\eeq{W11}
Here $m$ and $X$ are  antisymmetric $2\times 2$ matrices, and
$\l$ and $Z$ are symmetric $2\times 2$ matrices and
\beq
\G=M+\tr(ZX^{-1})^2 .
\eeq{111}
In ref. \cite{IS1}, both the
classical and quantum moduli spaces were described. Both classically and
quantum mechanically, the theory is generically in the Higgs/confinement
phase. The classical singularity at $X=Z=M=0$ is resolved quantum
mechanically into three singularities. We will reobtain this result in
detail.

We now want to find the vacua of the theory, namely, we should solve the
equations of motion $\d W_{1,1}/\d M=\d W_{1,1}/\d X=\d W_{1,1}/\d Z=0$,
which read:
\beq
\tm=2\L^{-3}(\pf X)\G,
\eeq{112}
\beq
m=R^{-1}(X^{-1}-8\G^{-1}X^{-1}(ZX^{-1})^2),
\eeq{113}
\beq
{1\over \sqrt{2}}\l=4R^{-1}\G^{-1}X^{-1}ZX^{-1},
\eeq{114}
where
\beq
R^{-1}=\L^{-3}(\pf X)\G^{2} .
\eeq{115}
Combining eqs. (\ref{113}) and (\ref{114}) we get
\beq
Xm+\sqrt{2}Z\l=R^{-1}I,
\eeq{116}
where $I$ is the $2\times 2$ identity matrix.
Equation (\ref{114}) gives
\beq
{1\over \sqrt{2}}Z\l={1\over 8}R\G(X\l)^2,
\eeq{117}
and using (\ref{117}), eq. (\ref{116}) reads:
\beq
Y^2+Y\n=2\G I,
\eeq{118}
where
\beq
\n={4\over \sqrt{2}}\l^{-1}m, \qquad Y={1\over \sqrt{2}}R\G X\l=4ZX^{-1},
\qquad \G=M+{1\over 16}\tr Y^2.
\eeq{119}
(The form of eqs. (\ref{111}), (\ref{113}), (\ref{114}),
(\ref{116})-(\ref{119}), as derived by variations from eq. (\ref{WfA}),
is $N_f$-independent for $N_A=1$ and will appear again in sections 6,7,8;
equations (\ref{W11}), (\ref{112}), (\ref{115}) contain, explicitly,
the $N_f$ dependence.)

Equations (\ref{119}), (\ref{115}) imply
\beq
\z_2\equiv \det Y=-{1\over 2}\tr Y^2={1\over 2}\L^6 \G^{-2} \det\l,
\qquad \tr Y=0 ,
\eeq{1111}
which implies that the characteristic polynomial of $Y$ is
\beq
Y^2+\z_2 I=0 .
\eeq{1112}
Using eqs. (\ref{118}) and (\ref{1112}) to eliminate $Y$ gives
\beq
\z_2\n^2+4\Big(\G+{1\over 2}\z_2\Big)^2 I=0 .
\eeq{1113}
The characteristic polynomial of $\n$ is
\beq
\n^2+\a_2 I=0, \qquad \a_2=\det\n, \qquad \tr\n=0 ,
\eeq{1114}
and with (\ref{1113}) and (\ref{1111}) we obtain
\beq
\a_2\z_2=4\Big(\G+{1\over 2}\z_2\Big)^2=4\L^6 \G^{-2}(\pf m)^2.
\eeq{1115}
{}From eqs. (\ref{119}) and (\ref{1111}) we read
\beq
\G=M-{1\over 8}\z_2 ,
\eeq{1116}
and, therefore, (\ref{1115}) becomes
\beq
\L^3\G^{-1}\pf m =\G +{1\over 2}\z_2=4M-3\G \implies 3\G^2-4M\G+\L^3 \pf m
=0 .
\eeq{1117}
Equations (\ref{1116}) and (\ref{1111}) imply
\beq
8(M-\G)={1\over 2}\L^6 \G^{-2}\det\l \implies \G^3-M\G^2+{1\over 16}\L^6
\det\l=0 .
\eeq{1118}
By combining eqs. (\ref{1117}) and (\ref{1118}) we find
\beq
x^3-Mx^2+{1\over 4}\L^3(\pf m)x-{1\over 64}\L^6\det\l=0,
\eeq{1119}
and
\beq
3x^2-2Mx+{1\over 4}\L^3\pf m=0,
\eeq{1120}
where
\beq
x\equiv {1\over 2}\G .
\eeq{1121}
Equations (\ref{1119}) and (\ref{1120}) are the singularity conditions of
an elliptic curve defined by
\beq
y^2=x^3+ax^2+bx+c,
\eeq{1122}
with
\beq
a=-M, \qquad b={\L^3\over 4} \pf m, \qquad c=-{\a\over 16},
\eeq{1123}
where
\beq
\a\equiv {\L^{2b_1}\over 2^{2N_f}} \det\l = {\L^6\over 4} \det\l .
\eeq{1124}

Solving $M$ in eq. (\ref{1120}), and eliminating $M$ in eq. (\ref{1119}) we
find
\beq
x^3-b x-2c=0,
\eeq{1125}
and
\beq
M={3\over 2}x+{b\over 2}x^{-1},
\eeq{1126}
Therefore, we find that $W_{1,1}$ (\ref{W11}) has three (branches of)
vacua, namely, the
three solutions for $M(x)$ in terms of the three solutions of the cubic
equation for $x$ (\ref{1125}) -- the singularities of the elliptic curve
(\ref{1122}), (\ref{1123}) -- and the solutions for $X$ and $Z$, given by
the other equations of motion; explicitly,
\beq
X={1\over \sqrt{2}}\tm Y\l^{-1}, \qquad Z={1\over 4\sqrt{2}}\tm Y^2\l^{-1},
\eeq{1126'}
where $Y$ is solved in terms of its invariants $\z_2$, given in eq.
(\ref{1111}), up to an $SU(2N_f)=SU(2)$ rotation, which is determined by
the by eq. (\ref{118}).

These three vacua are the vacua of the theory in the Higgs-confinement
phase. The phase transition points to the Coulomb branch are at $X=0$.
This happens iff the triplet superfield is massless, namely
\beq
X=0 \Leftrightarrow \tm=0.
\eeq{1128}
The coefficients $a,b,c$ of the ellipic curve and, in particular,  its
singularities, are independent of the value of
$X$ (namely, $\tm$)\footnote{
Note that $X=0 \implies \det Z=0$ in a way such that $x$ is finite.}
and, therefore, we conclude that the elliptic curve (\ref{1122}),
(\ref{1123}) defines the effective Abelian coupling,
$\tau(M,m,\l,\L)$, in the Coulomb
branch.\footnote{Note that the singularities of an elliptic curve, and its
behavior in asymptotic limits define it uniquely.}

Equation (\ref{1123}) generalizes \cite{IS1} the results of ref.
\cite{SW2} to {\em arbitrary} bare masses and Yukawa couplings. Indeed, in
the $N=2$ supersymmetric case (namely, when $\det\l=1$),
the result (\ref{1123}) coincides with the one obtained in ref. \cite{SW2}.
% All the symmetries and quantum numbers of the various parameters, as used
% in \cite{SW1,SW2}, are already embodied in the superpotential
% $W_{1,1}$ of eq.(\ref{W11}).

Finally, we should note that in eq. (\ref{1121}) we have identified a
physical meaning of the parameter $x$: it is the composite field
$\G/2$, where $\G$ is defined in eq. (\ref{111}).

\noindent
\section{$SU(2)$ with $N_A=1$, $N_f=2$ ($b_1=2$)}
\setcounter{equation}{0}

A supersymmetric $SU(2)$ gauge theory with one triplet and two flavors has a
superpotential (\ref{WfA}):
\beq
W_{2,1}=-2\L^{-1}(\pf X)^{1/2}\G+\tm M +{1\over 2}\tr\, mX +
{1\over \sqrt{2}} \tr\, \l Z.
\eeq{W21}
Here $m$ and $X$ are  antisymmetric $4\times 4$ matrices, and
$\l$ and $Z$ are symmetric $4\times 4$ matrices and
\beq
\G=M+\tr(ZX^{-1})^2 .
\eeq{211}

As in section 5, we now want to find the vacua of the theory, namely, we
should solve the equations of motion $\d W_{2,1}/\d M=\d W_{2,1}/\d X
=\d W_{2,1}/\d Z=0$, which read:
\beq
\tm=2\L^{-1}(\pf X)^{1/2},
\eeq{212}
\beq
m=R^{-1}(X^{-1}-8\G^{-1}X^{-1}(ZX^{-1})^2),
\eeq{213}
\beq
{1\over \sqrt{2}}\l=4R^{-1}\G^{-1}X^{-1}ZX^{-1},
\eeq{214}
where
\beq
R^{-1}=\L^{-1}(\pf X)^{1/2}\G.
\eeq{215}
Combining eqs. (\ref{213}) and (\ref{214}) we get
\beq
Xm+\sqrt{2}Z\l=R^{-1}I,
\eeq{216}
where $I$ is the $4\times 4$ identity matrix.
Equation (\ref{214}) gives
\beq
{1\over \sqrt{2}}Z\l={1\over 8}R\G(X\l)^2,
\eeq{217}
and using (\ref{217}), eq. (\ref{216}) reads:
\beq
Y^2+Y\n=2\G I,
\eeq{218}
where
\beq
\n={4\over \sqrt{2}}\l^{-1}m, \qquad Y={1\over \sqrt{2}}R\G X\l=4ZX^{-1},
\qquad \G=M+{1\over 16}\tr Y^2.
\eeq{219}
Equation (\ref{218}) implies, in particular,
\beq
[Y,\n]=0 ,
\eeq{2110}
and from (\ref{219}), (\ref{215}) we get
\beq
\z_2\equiv -{1\over 2}\tr Y^2=8(M-\G),\qquad
\z_4\equiv \det Y={1\over 4}\L^4 \det\l, \qquad \tr Y=\tr Y^3=0 ,
\eeq{2111}
thus the characteristic polynomial of $Y$ is
\beq
Y^4+\z_2 Y^2+\z_4 I=0 .
\eeq{2112}
Using eqs. (\ref{218}), (\ref{2110}) to eliminate $Y^4$ in (\ref{2112}),
we get
\beq
(\n^2+\z_2)Y^2-4\G\n Y+(4\G^2+\z_4)I=0 .
\eeq{2113}
Now, using eqs. (\ref{218}), (\ref{2110})
and (\ref{2113}) to eliminate $Y$, we get
\beq
\z_4\n^4+(4\G^2\z_2+8\G\z_4+\z_2\z_4)\n^2+(4\G^2+2\G\z_2+\z_4)^2 I=0 .
\eeq{2113'}
The characteristic polynomial of $\n$ is
\beq
\n^4+\a_2\n^2+\a_4 I=0, \qquad \a_4=\det\n,\quad \a_2=-{1\over 2}\tr \n^2,
\qquad \tr\n=\tr\n^3=0 ,
\eeq{2114}
and with (\ref{2113'}) and (\ref{2111}) we obtain
\beq
\a_2\z_4=4\G^2\z_2+8\G\z_4+\z_2\z_4 = 8(M-\G)(4\G^2+\z_4)+8\G\z_4,
\eeq{2115}
and
\beq
(\a_4\z_4)^{1/2}=4\G^2+2\G\z_2+\z_4 = 16\G(M-\G)+4\G^2+\z_4.
\eeq{2115'}
{}From eqs. (\ref{2115}), (\ref{2115'}) we find
\beq
x^3-Mx^2+{(\a_4\z_4)^{1/2}-\z_4\over 16}x-{1\over 128}(\a_2-8M)\z_4=0,
\eeq{2116}
and
\beq
3x^2-2Mx+{(\a_4\z_4)^{1/2}-\z_4\over 16}=0,
\eeq{2117}
where
\beq
x\equiv {1\over 2}\G .
\eeq{2118}
Equations (\ref{2116}) and (\ref{2117}) are the singularity conditions of
an elliptic curve defined by
\beq
y^2=x^3+ax^2+bx+c,
\eeq{2119}
with
\ber
a&=& -M, \qquad b\,\,=\,\, -{\a\over 4}+{\L^2\over 4}\pf m , \nonumber\\
c&=& {\a\over 8}\Big(2M+\tr(\m^2)\Big),
\eer{abc}
where
\beq
\a\equiv {\L^{2b_1}\over 2^{2N_f}} \det\l = {\L^4\over 16} \det\l, \qquad
\m=\l^{-1}m.
\eeq{aL2}
Here we used the explicit expressions for $\z_4,\a_4,\a_2$ in terms of
$\L,\l,m$ (see eqs. (\ref{219}), (\ref{2111}), (\ref{2114})).

Solving $M$ in eq. (\ref{2117}), and eliminating $M$ in eq. (\ref{2116}) we
find
\beq
-x^4+\Big(b+{3\over 4}\a\Big)x^2+{\a\over 4}\tr(\m^2)x+{\a b\over 4}=0,
\eeq{2120}
and
\beq
M={3\over 2}x+{b\over 2}x^{-1},
\eeq{2121}
Therefore, we find that $W_{2,1}$ (\ref{W21}) has four (branches of) vacua,
namely, the
four solutions for $M(x)$ in terms of the four solutions of the quartic
equation for $x$ (\ref{2120}) -- the singularities of the elliptic curve
(\ref{2119}), (\ref{abc}) -- and the solutions for $X$ and $Z$, given by
the other equations of motion; explicitly,
\beq
X={1\over \sqrt{2}}\tm Y\l^{-1}, \qquad Z={1\over 4\sqrt{2}}\tm Y^2\l^{-1},
\eeq{2126'}
where $Y$ is solved in terms of its invariants $\z_2, \z_4$, given in eq.
(\ref{2111}), up to an $SU(2N_f)=SU(4)$ rotation, which is determined by
eq. (\ref{218}).

These four vacua are the vacua of the theory in the Higgs-confinement phase.
The phase transition points to the Coulomb branch are at $X=0$.
This may happen if the triplet superfield is massless, namely, $\tm=0$.
The coefficients $a,b,c$ of the ellipic curve and, in particular, its
singularities, are independent of the value of $X$ (namely, $\tm$)
and, therefore, we conclude that the elliptic curve (\ref{2119}),
(\ref{abc}) defines the effective Abelian coupling,
$\tau(M,m,\l,\L)$, in the Coulomb branch.

Equation (\ref{abc}) generalizes the result of ref. \cite{SW2} to {\em
arbitrary} bare masses and Yukawa couplings.
Indeed, in the $N=2$ supersymmetric case (namely, when $\l={\rm
diag(\l_1,\l_2)}$, where $\l_1,\l_2$ are $2\times 2$ matrices with
$\det\l_1=\det\l_2=1$, and $m={\rm diag}(m_1\e,m_2\e)$, where $\e$ is the
standard $2\times 2$ constant antisymmetric matrix), the result (\ref{abc})
coincides with the one obtained in ref. \cite{SW2}.
% All the symmetries and quantum numbers of the various parameters, as used
% in \cite{SW1,SW2}, are already embodied in the superpotential
% $W_{2,1}$ of eq. (\ref{W21}).

Finally, as in the $N_f=1$ case,
we should note that in eq. (\ref{2118}) we have identified a
physical meaning of the parameter $x$.

\noindent
\section{$SU(2)$ with $N_A=1$, $N_f=3$ ($b_1=1$)}
\setcounter{equation}{0}

A supersymmetric $SU(2)$ gauge theory with one triplet and three flavors
has a superpotential (\ref{WfA}):
\beq
W_{3,1}=-3\L^{-{1/3}}(\pf X)^{1/3}\G^{2/3}+\tm M +{1\over 2}\tr\, mX +
{1\over \sqrt{2}} \tr\, \l Z.
\eeq{W31}
Here $m$ and $X$ are  antisymmetric $6\times 6$ matrices, and
$\l$ and $Z$ are symmetric $6\times 6$ matrices and
\beq
\G=M+\tr(ZX^{-1})^2 .
\eeq{311}

As in sections 5 and 6, we now want to find the vacua of the theory,
namely, we should solve the equations of motion
$\d W_{3,1}/\d M=\d W_{3,1}/\d X=\d W_{3,1}/\d Z=0$, which read:
\beq
\tm=2\L^{-1/3}(\pf X)^{1/3}\G^{-1/3},
\eeq{312}
\beq
m=R^{-1}(X^{-1}-8\G^{-1}X^{-1}(ZX^{-1})^2),
\eeq{313}
\beq
{1\over \sqrt{2}}\l=4R^{-1}\G^{-1}X^{-1}ZX^{-1},
\eeq{314}
where
\beq
R^{-1}=\L^{-1/3}(\pf X)^{1/3}\G^{2/3}.
\eeq{315}
Combining eqs. (\ref{313}) and (\ref{314}) we get
\beq
Xm+\sqrt{2}Z\l=R^{-1}I,
\eeq{316}
where $I$ is the $6\times 6$ identity matrix.
Equation (\ref{314}) gives
\beq
{1\over \sqrt{2}}Z\l={1\over 8}R\G(X\l)^2,
\eeq{317}
and using (\ref{317}), eq. (\ref{316}) reads:
\beq
Y^2+Y\n=2\G I,
\eeq{318}
where
\beq
\n={4\over \sqrt{2}}\l^{-1}m, \qquad Y={1\over \sqrt{2}}R\G X\l=4ZX^{-1},
\qquad \G=M+{1\over 16}\tr Y^2,
\eeq{319}
and (\ref{318}) implies, in particular,
\beq
[Y,\n]=0 .
\eeq{3110}
{}From eqs. (\ref{319}), (\ref{315}) we get
\ber
\z_2\equiv -{1\over 2}\tr Y^2=8(M-\G),\, & & \,
\z_4\equiv {1\over 2}\z_2^2-{1\over 4}\tr Y^4 , \nonumber\\
\z_6\equiv \det Y={1\over 8}\G^2\L^2 \det\l, \, & & \,
\tr Y=\tr Y^3=\tr Y^5=0 ,
\eer{3111}
determining the characteristic polynomial of $Y$ as
\beq
Y^6+\z_2 Y^4+\z_4 Y^2+ \z_6 I=0 .
\eeq{3112}
Using eqs. (\ref{318}), (\ref{3110}) to eliminate
$Y^4$ and $Y^6$ in (\ref{3112}), we get
\ber
[\n^4+(\z_2+4\G)\n^2+\z_4+4\G^2]Y^2&-&[4\G\n^3+\n(8\G^2+4\z_2\G)]Y
\nonumber\\ &+& [\z_6+4\z_2\G^2+4\G^2\n^2]I=0 . \nonumber\\
\eer{3113}
Now, using eqs. (\ref{318}), (\ref{3110})
and (\ref{3113}) to eliminate $Y$, we get
\ber
\z_6\n^6&+&(4\G^2\z_4+12\G\z_6+\z_2\z_6)\n^4 \nonumber\\&+&
(16\G^4\z_2+32\G^3\z_4+4\G^2\z_2\z_4+36\G^2\z_6+8\G\z_2\z_6+\z_4\z_6)\n^2
\nonumber\\&+&(8\G^3+4\G^2\z_2+2\G\z_4+\z_6)^2 I=0 .
\eer{3113'}
The characteristic polynomial of $\n$ is
\ber
\n^6+\a_2\n^4+\a_4\n^2+\a_6 I=0, \,& &\,\a_2=-{1\over 2}\tr \n^2,\quad
\a_4={1\over 2}\a_2^2-{1\over 4}\tr\n^4, \nonumber\\
\a_6=\det\n,\,\,& &\,\, \tr\n=\tr\n^3=\tr\n^5=0 ,
\eer{3114}
and with (\ref{3113'}) and (\ref{3111}) we obtain
\beq
\a_2\z_6=4\G^2\z_4+12\G\z_6+\z_2\z_6 = 4\G^2\z_4+4(2M+\G)\z_6,
\eeq{3115}
\ber
\a_4\z_6&=&16\G^4\z_2+32\G^3\z_4+4\G^2\z_2\z_4+36\G^2\z_6+8\G\z_2\z_6+\z_4\z_6
\nonumber\\&=&
8(M-\G)(16\G^4+4\G^2\z_4+8\G\z_6)+32\G^3\z_4+36\G^2\z_6+\z_4\z_6,\nonumber\\
\eer{3115'}
and
\beq
(\a_6\z_6)^{1/2}=8\G^3+4\G^2\z_2+2\G\z_4+\z_6=8\G^3+32\G^2(M-\G)+2\G\z_4+\z_6
{}.
\eeq{3115''}
Eliminating $\z_4$ from eqs. (\ref{3115}), (\ref{3115'}), (\ref{3115''}),
and shifting $M\to M-\L^2\det\l/256$ we find
\beq
x^3+ax^2+bx+c=0 ,
\eeq{3116}
and
\beq
3x^2+2ax+b=0 .
\eeq{3117}
Here
\beq
x\equiv {1\over 2}\G+{\L^2\over 128}\det\l ,
\eeq{3118}
and
\ber
a&=& -M-\a ,   \nonumber\\
b&=& 2\a  M + {\a\over 2}\tr(\m^2) + {\L\over 4} \pf m ,
\nonumber\\
c&=&{\a\over 8}\Big( -8M^2-4M \tr(\m^2) - [\tr(\m^2)]^2 + 2\tr(\m^4)
\Big) ,
\eer{abc3}
where
\beq
\a\equiv {\L^{2b_1}\over 2^{2N_f}}\det\l={\L^2\over 64}\det \l,
\qquad \m=\l^{-1}m.
\eeq{am}
Here we used the explicit expressions for $\z_6,\a_6,\a_4,\a_2$ in terms
of $\L,\l,m$ (see eqs. (\ref{319}), (\ref{3111}), (\ref{3114})).
Equations (\ref{3116}) and (\ref{3117}) are the singularity conditions of
an elliptic curve defined by
\beq
y^2=x^3+ax^2+bx+c,
\eeq{3119}
with coefficients $a,b,c$ given in eqs. (\ref{abc3}), (\ref{am}).

Solving $M$ in eq. (\ref{3117}), and eliminating $M$ in eq. (\ref{3116})
one finds a degree 5 polynomial equation in $x$, $p_5(x)=0$, and an equation
for $M(x)$.
Therefore, we find that $W_{3,1}$ (\ref{W31}) has five (branches of) vacua,
namely, the
five solutions for $M(x)$ in terms of the five solutions of the fifth order
equation for $x$ -- the singularities of the elliptic curve
(\ref{3119}), (\ref{abc3}) -- and the solutions for $X$ and $Z$, given by
the other equations of motion; explicitly,
\beq
X={1\over \sqrt{2}}\tm Y\l^{-1}, \qquad Z={1\over 4\sqrt{2}}\tm Y^2\l^{-1},
\eeq{3126'}
where $Y$ is solved in terms of its invariants $\z_2, \z_4, \z_6$, given
in eq. (\ref{3111}) (to find $\z_4$ we use eq. (\ref{3115})),
up to an $SU(2N_f)=SU(6)$ rotation, determined by eq. (\ref{318}).

These five vacua are the vacua of the theory in the Higgs-confinement
phase. The phase transition points to the Coulomb branch are at $X=0$.
This may happen if the triplet superfield is massless, namely, $\tm=0$.
The coefficients $a,b,c$ of the ellipic curve and, in particular, its
singularities, are independent of the value of $X$ (namely, $\tm$)
and, therefore, we conclude that the elliptic curve (\ref{3119}),
(\ref{abc3}) defines the effective Abelian coupling,
$\tau(M,m,\l,\L)$, in the Coulomb branch.

Equation (\ref{abc3}) generalizes the result of ref. \cite{SW2} to {\em
arbitrary} bare masses and Yukawa couplings.
Indeed, in the $N=2$ supersymmetric case (namely, when $\l={\rm
diag(\l_1,\l_2,\l_3)}$, where $\l_1,\l_2,\l_3$ are $2\times 2$ matrices with
$\det\l_1=\det\l_2=\det\l_3=1$, and $m={\rm diag}(m_1\e,m_2\e,m_3\e)$,
where $\e$ is the standard $2\times 2$ constant antisymmetric matrix), the
result (\ref{abc3}) coincides with the one obtained in ref. \cite{SW2}.
% All the symmetries and quantum numbers of the various parameters, as used
% in \cite{SW1,SW2}, are already embodied in the superpotential $W_{3,1}$
% of eq. (\ref{W31}).

Finally, as before, we should note that in eq. (\ref{3118}) we have
identified a physical meaning of the parameter $x$. Unlike the $N_f=1$
and $N_f=2$ cases, for $N_f=3$, $x$ is identified with $\G/2$ only up to a
shift by $\a/2$, where $\G$ and $\a$ are given in eqs. (\ref{311}) and
(\ref{am}), respectively.

\noindent
\section{$SU(2)$ with $N_A=1$, $N_f=4$ ($b_1=0$)}
\setcounter{equation}{0}

A supersymmetric $SU(2)$ gauge theory with one triplet and four flavors has a
vanishing one-loop beta-function and, therefore, will possess extra structure.
It has a
superpotential (\ref{WfA}):~\footnote{$b_1=0$ and, as noted before,
``$\L^{-b_1}$'' in (\ref{W41}) should be replaced by a function
of $\tau_0={\th_0\over\pi}+{8\pi i\over g_0^2}$ (the non-Abelian gauge
coupling constant) and $\det \l$; the issue is addressed in this section.}
\beq
W_{4,1}=-4\L^{-b_1/4}(\pf X)^{1/4}\G^{1/2}+\tm M +{1\over 2}\tr\, mX +
{1\over \sqrt{2}} \tr\, \l Z.
\eeq{W41}
Here $m$ and $X$ are  antisymmetric $8\times 8$ matrices, and
$\l$ and $Z$ are symmetric $8\times 8$ matrices and
\beq
\G=M+\tr(ZX^{-1})^2 .
\eeq{411}

As in sections 5, 6 and 7, we now want to find the vacua of the
theory, namely, we should solve the equations of motion
$\d W_{4,1}/\d M=\d W_{4,1}/\d X=\d W_{4,1}/\d Z=0$, which read:
\beq
\tm=2\L^{-b_1/4}(\pf X)^{1/4}\G^{-1/2},
\eeq{412}
\beq
m=R^{-1}(X^{-1}-8\G^{-1}X^{-1}(ZX^{-1})^2),
\eeq{413}
\beq
{1\over \sqrt{2}}\l=4R^{-1}\G^{-1}X^{-1}ZX^{-1},
\eeq{414}
where
\beq
R^{-1}=\L^{-b_1/4}(\pf X)^{1/4}\G^{1/2}.
\eeq{415}
Combining eqs. (\ref{413}) and (\ref{414}) we get
\beq
Xm+\sqrt{2}Z\l=R^{-1}I,
\eeq{416}
where $I$ is the $8\times 8$ identity matrix.
Equation (\ref{414}) gives
\beq
{1\over \sqrt{2}}Z\l={1\over 8}R\G(X\l)^2,
\eeq{417}
and using (\ref{417}), eq. (\ref{416}) reads:
\beq
Y^2+Y\n=2\G I,
\eeq{418}
where
\beq
\n={4\over \sqrt{2}}\l^{-1}m, \qquad Y={1\over \sqrt{2}}R\G X\l=4ZX^{-1},
\qquad \G=M+{1\over 16}\tr Y^2.
\eeq{419}
Again, eq. (\ref{418}) implies, in particular,
\beq
[Y,\n]=0,
\eeq{4110}
and eqs. (\ref{419}), (\ref{415}) imply
\ber
\z_2\equiv -{1\over 2}\tr Y^2=8(M-\G),\, & & \,
\z_4\equiv {1\over 2}\z_2^2-{1\over 4}\tr Y^4 , \nonumber\\
\z_6\equiv \z_2\Big(\z_4-{1\over 3}\z_2^2\Big)-{1\over 6}\tr Y^6 \, & & \,
\z_8\equiv \det Y={1\over 16}\G^4\L^{2b_1} \det\l=16\G^4\a(\t_0),
\nonumber\\ \, & & \, \tr Y=\tr Y^3=\tr Y^5=\tr Y^7=0 ,
\eer{4111}
Here we replaced $\L^{b_1}$ with a function of $\l$ and the non-Abelian
gauge coupling, $\t_0$, in a way consistent with the global
symmetries:~\footnote{
$(\det\l)^{-1/2}$ has the correct quantum numbers needed for the matching
condition, $16\a^{1/2}(\det\l)^{-1/2}\tm^2=\L^2_{N_f=4,N_A=0}$,
where $\a(\tau_0)$ is dimensionless,
and has zero $U(1)_R\times U(1)_Q\times U(1)_{\P}$ quantum numbers.}
\beq
\L^{b_1}\equiv 16\a(\t_0)^{1/2}(\det\l)^{-1/2},
\eeq{4111'}
where $\a(\t_0)$ will be determined later.
Equation (\ref{4111}) implies that the characteristic polynomial of $Y$ is
\beq
Y^8+\z_2 Y^6+\z_4 Y^4+ \z_6 Y^2 + \z_8 I=0 .
\eeq{4112}
Using eqs. (\ref{418}), (\ref{4110}) to eliminate
$Y^4$, $Y^6$ and $Y^8$ in (\ref{4112}), we get
\ber
& &[\n^6+(8\G+\z_2)\n^4+(16\G^2+4\G\z_2+\z_4)\n^2+4\G^2\z_2+\z_6]Y^2
\nonumber\\
&-&[4\G\n^5+(24\G^2+4\G\z_2)\n^3+(32\G^3+8\G^2\z_2+4\G\z_4)\n]Y
\nonumber\\
&+&[4\G^2\n^4+(16\G^3+4\G^2\z_2)\n^2+16\G^4+4\G^2\z_4+\z_8]I=0 .
\eer{4113}
Now, using eqs. (\ref{418}), (\ref{4110})
and (\ref{4113}) to eliminate $Y$, we get
\ber
\z_8\n^8&+&(4\G^2\z_6+16\G\z_8+\z_2\z_8)\n^6 \nonumber\\&+&
(16\G^4\z_4+48\G^3\z_6+80\G^2\z_8+4\G^2\z_2\z_6+12\G\z_2\z_8+\z_4\z_8)\n^4
\nonumber\\&+&(64\G^6\z_2+128\G^5\z_4+144\G^4\z_6+16\G^4\z_2\z_4+128\G^3\z_8
+32\G^3\z_2\z_6\nonumber\\
\,& &\,\, +36\G^2\z_2\z_8+4\G^2\z_4\z_6+8\G\z_4\z_8+\z_6\z_8)\n^2
\nonumber\\&+&(16\G^4+8\G^3\z_2+4\G^2\z_4+2\G\z_6+\z_8)^2 I=0 .
\eer{4113'}
The characteristic polynomial of $\n$ is
\ber
\n^8+\a_2\n^6+\a_4\n^4+\a_6\n^2+\a_8I=0, \,& &\,\a_2=-{1\over 2}\tr \n^2,\quad
\a_4={1\over 2}\a_2^2-{1\over 4}\tr\n^4, \nonumber\\
\a_6=\a_2\Big(\a_4-{1\over 3}\a_2^2\Big)-{1\over 6}\tr\n^6, \,& &\,
\a_8=\det\n,\nonumber\\ & &\,\, \tr\n=\tr\n^3=\tr\n^5=\tr\n^7=0 ,
\nonumber\\
\eer{4114}
and with (\ref{4113'}) and (\ref{4111}) we obtain
\beq
\a_2\z_8=4\G^2\z_6+16\G\z_8+\z_2\z_8 ,
\eeq{4115}
\beq
\a_4\z_8=16\G^4\z_4+48\G^3\z_6+80\G^2\z_8+4\G^2\z_2\z_6+12\G\z_2\z_8+\z_4\z_8
,
\eeq{4115'}
\ber
\a_6\z_8&=&64\G^6\z_2+128\G^5\z_4+144\G^4\z_6+16\G^4\z_2\z_4+128\G^3\z_8
\nonumber\\&+&
32\G^3\z_2\z_6+36\G^2\z_2\z_8+4\G^2\z_4\z_6+8\G\z_4\z_8+\z_6\z_8 ,
\eer{4115''}
and
\beq
(\a_8\z_8)^{1/2}=16\G^4+8\G^3\z_2+4\G^2\z_4+2\G\z_6+\z_8 .
\eeq{4115'''}
Inserting in (\ref{4115})-(\ref{4115'''}) the explicit expressions of
$\z_8$ and $\a_8$ in eqs. (\ref{4111}) and (\ref{4114}), respectively,
then eliminating $\z_6$ and $\z_4$, and after some algebra we find
\ber
& &256(\a-1)^2\G^3+64(\a-1)[8M(\a+1)-\a_2\a]\G^2\nonumber\\
&+&16[(\a+1)\L^{b_1}\pf m-4\a(\a_4+8M(8M-\a_2))]\G \nonumber\\
&+&16[\a_2\a+8M(1-\a)]\Big[{-\a\over (\a-1)^2} \L^{b_1}\pf m
+{\a(\a+1)\over (\a-1)^2}[\a_4+8M(8M-\a_2)]\Big]\nonumber\\
&+&{8\a_2\a\over (\a-1)^2}[(\a+1)\L^{b_1}\pf m -4\a(\a_4+8M(8M-\a_2))]
-16\a_6\a=0 ,
\eer{4116'}
and
\ber
&&768(\a-1)^2\G^2+128(\a-1)[8M(\a+1)-\a_2\a]\G\nonumber\\
&+&16(\a+1)\L^{b_1}\pf m -64\a[\a_4+8M(8M-\a_2)]=0 .
\eer{4117'}
Shifting and rescaling $M$:
$M\to \b^2\Big[M-{\a \over \b^2(\a-1)}\tr\m^2\Big]$, were $\b\equiv
\b(\t_0)$ is to be determined,  we find that (\ref{4116'}) and
(\ref{4117'}) become
\beq
x^3+ax^2+bx+c=0 ,
\eeq{4116}
and
\beq
3x^2+2ax+b=0 .
\eeq{4117}
Here
\beq
x\equiv {1\over \b^4}\Big[\G-{4\a \over (\a-1)^2}\tr\m^2\Big] ,
\qquad \m=\l^{-1}m,
\eeq{4118}
and
\ber
a&=&{1\over \b^2}\Big\{
2{\a+1\over \a-1}M+{8\over \b^2}{\a\over (\a-1)^2}\tr(\m^2)\Big\}, \nonumber\\
b&=&{1\over \b^4}\Big\{
-16{\a\over (\a-1)^2} M^2 + {32\over \b^2}{\a(\a+1)\over (\a-1)^3}M\tr(\m^2)
\nonumber\\
&-&{8\over \b^4}{\a\over (\a-1)^2}\Big[(\tr (\m^2))^2-2\tr(\m^4)\Big]+
{4\over \b^4}{(\a+1)\L^{b_1}\over (\a-1)^2} \pf m\Big\} , \nonumber\\
c&=& {1\over \b^6}\Big\{
-32{\a(\a+1)\over (\a-1)^3}M^3+{32\over \b^2}{\a(\a+1)^2\over
(\a-1)^4}M^2\tr(\m^2)\nonumber\\
&+&M\Big[-{16\over \b^4}{\a(\a+1)\over (\a-1)^3}
\Big((\tr(\m^2))^2-2\tr(\m^4)\Big) + {32\over \b^4}{\a\L^{b_1}\over (\a-1)^3}
\pf m \Big]\nonumber\\
&-&{32\over \b^6}{\a\over (\a-1)^2}\Big[\tr(\m^2)\tr(\m^4)-{1\over
6}(\tr(\m^2))^3-{4\over 3}\tr(\m^6)\Big]\Big\}.
\eer{abc4}

Equations (\ref{4116}) and (\ref{4117}) are the singularity conditions of
an elliptic curve defined by
\beq
y^2=x^3+ax^2+bx+c,
\eeq{4119}
with coefficients $a,b,c$ given in eqs. (\ref{abc4}).
The singularity condition for the elliptic curve is equivalent to the
vanishing condition of the discriminant:
\beq
\D=4a^3c-a^2b^2-18abc+4b^3+27c^2=0 ,
\eeq{4120}
and, moreover, from eqs. (\ref{4116}), (\ref{4117}) we can get
\beq
x={ab-9c\over 6b-2a^2}.
\eeq{4120'}
$\D (M)$ is a polynomial in $M$ of degree 6 and, therefore, there are six
(branches of) vacua ($x\equiv x(M)$ is given by eq. (\ref{4120'}), and the
solutions for $X$ and $Z$ are given by the other equations of motion,
as was done in sections 5,6,7 for $N_f<4$).
These are the vacua of the theory in the Higgs/confinement phase.
The phase transition points to the Coulomb branch are at $X=0$.
This may happen if the triplet superfield is massless, namely, $\tm=0$.
The coefficients $a,b,c$ of the ellipic curve and, in particular, its
singularities, are independent of the value of $X$ (namely, $\tm$)
and, therefore, we conclude that the elliptic curve (\ref{4119}),
(\ref{abc4}) defines the effective Abelian coupling,
$\tau(M,m,\l,\L)$, in the Coulomb branch.

We should now determine $\a$ and $\b$. They
are functions of $\tau_0$, the non-Abelian
gauge coupling constant; comparison with ref. \cite{SW2} gives~\footnote{
To compare eq. (\ref{abc4}) with the $N=2$ supersymmetric case in ref.
\cite{SW2} we need to take $m={\rm diag}(m_1\e,m_2\e,m_3\e,m_4\e)$ and
$\l={\rm diag}(\l_1,\l_2,\l_3,\l_4)$, where $\l_I$, $I=1,2,3,4$, are
$2\times 2$ matrices with $\det(\l_I)=1$. In this case,
$\tr(\m^2)=-2\sum_{I=1}^4 m_I^2$,
$(\tr(\m^2))^2-2\tr(\m^4)=8\sum_{I<J}m_I^2 m_J^2$,
$\tr(\m^2)\tr(\m^4)-{1\over 6}(\tr(\m^2))^3-{4\over
3}\tr(\m^6)=8\sum_{I<J<K} m_I^2 m_J^2 m_K^2$.}
\beq
\a(\tau_0)\equiv {``\L^{2b_1}"\over 2^{2N_f}} \det\l
=\left({\th_2^2-\th_3^2\over \th_2^2+\th_3^2}\right)^2, \qquad
\b(\tau_0)={\sqrt{2}\over \th_2\th_3},
\eeq{g2}
where
\beq
\th_2(\tau_0)=\sum_{n\in Z}(-1)^n e^{\pi i \tau_0 n^2}, \qquad
\th_3(\tau_0)=\sum_{n\in Z}e^{\pi i \tau_0 n^2}, \qquad
\tau_0={\th_0\over \pi}+{8\pi i\over g_0^2}.
\eeq{gth}
Equation (\ref{abc4}) generalizes the result of ref. \cite{SW2} to {\em
arbitrary} bare masses and Yukawa couplings.
As in the other cases,
all the symmetries and quantum numbers of the various parameters, as used
in \cite{SW1,SW2}, are already embodied in the superpotential $W_{4,1}$ of eq.
(\ref{W41}).

The $S$-duality symmetry is valid in the $N_A=1, N_f=4$ theories for {\em
arbitrary} $\l, m$,  similar to the $SL(2,Z)$ invariance in the presence of
masses discussed in ref. \cite{SW2}. The $SL(2,Z)$ transformations map
$\tau_0$ to $(a\tau_0+b)(c\tau_0+d)^{-1}$, $a,b,c,d\in Z$,
$ad-bc=1$ \cite{CR}. Combined
with triality (which acts on $\m$), it leaves the elliptic curve invariant.

Finally, we should note again that in eq. (\ref{4118}) we have identified a
physical meaning of the parameter $x$. Unlike the $N_f<4$ cases, for
$N_f=4$, $x$ is identified with $\G$ only up to a shift by a $\m$-dependent
and a $\tau_0$-dependent function (which vanish at $m=0$), and a rescaling by
a $\tau_0$-dependent function.

\noindent
\section{$SU(2)$ with $N_A=2$, $N_f=1$ ($b_1=1$)}
\setcounter{equation}{0}

The $N=1$ supersymmetric $SU(2)$ gauge theory with two triplets ($N_A=2$,
$N_f=0$) was discussed in section 4. In this section, we consider a
supersymmetric $SU(2)$ gauge theory with two triplets and one flavor.
The superpotential (\ref{WfA}) is
\beq
W_{1,2}=-3\L^{-1/3}(\pf X)^{1/3}(\det\G)^{2/3}+\tr_{N_A}\tm M
+{1\over 2}\tr_{2N_f} mX + {1\over \sqrt{2}} \tr_{2N_f} \l^{\a} Z_{\a}.
\eeq{W12}
Here $m$ and $X$ are  antisymmetric $2\times 2$ matrices,
$\l^{\a}$ and $Z_{\a}$ are symmetric $2\times 2$ matrices, $\a=1,2$,
$\tm$, $M$ are $2\times 2$ symmetric matrices and
\beq
\G_{\a\b}=M_{\a\b}+\tr(Z_{\a}X^{-1}Z_{\b}X^{-1}) .
\eeq{121}

We now want to find the vacua of the theory, namely, we should solve the
equations of motion
$\d W_{1,2}/\d M_{\a\b}=\d W_{1,2}/\d X=\d W_{1,2}/\d Z_{\a}=0$. The
procedure is similar to the $N_A=N_f=1$ case, with the additional
complication induced by the matrix structure of $\G$; the equations of motion
read:
\beq
\tm_{\a\b}=2R^{-1}(\G^{-1})^{\a\b} ,
\eeq{122}
\beq
m=R^{-1}(X^{-1}-8(\G^{-1})^{\a\b}X^{-1}Z_{\a}X^{-1}Z_{\b}X^{-1}),
\eeq{123}
\beq
{1\over \sqrt{2}}\l^{\a}=4R^{-1}(\G^{-1})^{\a\b}X^{-1}Z_{\b}X^{-1},
\eeq{124}
where
\beq
R^{-1}=\L^{-1/3}(\pf X)^{1/3}(\det\G)^{2/3} .
\eeq{125}
Combining eqs. (\ref{123}) and (\ref{124}) we get
\beq
Xm+\sqrt{2}Z_{\a}\l^{\a}=R^{-1}I,
\eeq{126}
where $I$ is the $2\times 2$ identity matrix.

Inserting (\ref{124}) in (\ref{121}), and using $X=\e\pf X$, where $\e$ is
the standard antisymmetric $2\times 2$ matrix, we obtain
\beq
M=\G-\Big({R\pf X\over 4}\Big)^2 \G S \G ,
\eeq{127}
where
\beq
S^{\a\b}={1\over 2}\tr(\e\l^{\a}\e\l^{\b}) .
\eeq{128}
Using eq. (\ref{122}) we find
\beq
M_{\a\b}={1\over 2}R(\det\G)\m_{\a\b}-\Big({R^2\pf X \det\G\over 8}\Big)^2
\hat{S}_{\a\b},
\eeq{129}
where
\beq
\m_{11}=\tm_{22}, \qquad \m_{22}=\tm_{11}, \qquad \m_{12}=\m_{21}=-\tm_{12},
\eeq{1210}
and
\ber
\hat{S}_{11}&=&S^{11}\tm_{22}^2+S^{22}\tm_{12}^2-2S^{12}\tm_{12}\tm_{22},
\nonumber\\
\hat{S}_{12}&=&-S^{11}\tm_{12}\tm_{22}-S^{22}\tm_{12}\tm_{11}+
S^{12}(\tm_{11}\tm_{22}+\tm_{12}^2),
\nonumber\\
\hat{S}_{22}&=&S^{11}\tm_{12}^2+S^{22}\tm_{11}^2-2S^{12}\tm_{12}\tm_{11}=
-\det (\tm_{12}\l^1-\tm_{11}\l^2) . \nonumber\\
\eer{1211}

{}From eqs. (\ref{124}), (\ref{126}), (\ref{128}) we get
\beq
{1\over 4}R^2(\pf X)^2 \G_{\a\b}S^{\a\b} = 1-R\pf (mX),
\eeq{1212}
and using eq. (\ref{122}) we obtain
\beq
{1\over 8}R^3(\pf X)^2 (\det\G)\hat{S}_{22}+{1\over 2}R(\pf X)^2 S^{11}
=\tm_{11}[1-R\pf(mX)] .
\eeq{1213}
Using eq. (\ref{125}), and after some algebra, eqs. (\ref{129}) and
(\ref{1213}) read:
\ber
M_{11}&=&{1\over 2}R(\det\G)\tm_{22}-\Big({R^2\pf X \det\G\over 8}\Big)^2
\hat{S}_{11},\nonumber\\
M_{12}&=&-{1\over 2}R(\det\G)\tm_{12}-\Big({R^2\pf X \det\G\over 8}\Big)^2
\hat{S}_{12},
\eer{1214}
\ber
&&x^3-M_{22}x^2+{1\over 16}\L\tm_{11}^2(\pf m)x \nonumber\\
&+&\Big({\L \tm_{11}\over 32}\Big)^2\hat{S}_{22}
-{1\over 512} \L^2\tm_{11}^2\det\tm\det\l^1 =0,
\eer{1215}
\beq
3x^2-2M_{22}x+{1\over 16}\L\tm_{11}^2\pf m
-\Big({1\over 512} \L^2\tm_{11}^2\det\tm\det\l^1\Big){1\over x}=0 ,
\eeq{1216}
where
\beq
x\equiv {1\over 4}R(\det\G)\tm_{11}={\tm_{11}\over 2(\det
\tm)^{1/2}}(\det\G)^{1/2} .
\eeq{1217}

The solutions of eqs. (\ref{1214}), (\ref{1215}) and (\ref{1216})
determine the vacua of the theory; there are three (branches of) vacua.
(Integrating out the doublets, namely, taking $m\to\infty$ keeping
$\L \pf m$ fixed, one is left with two vacua). We now want to consider
the phase transition points from the confinement branch to the Coulomb branch.
This happens at vacua where $\langle\det M\rangle=0$, namely, when
$\det\tm=0$. Explicitly, without loss of generality, we study the
case when the triplet $\P_2$ is massless:
\beq
\tm_{22}=\tm_{12}=0 ,
\eeq{1218}
so inserting (\ref{1218}) in eq. (\ref{1214}) implies
\beq
M_{11}=M_{12}=0 \implies \det\, M=0 .
\eeq{1219}
Inserting (\ref{1218}) in eqs. (\ref{1215}) and (\ref{1216}) turn them into
\beq
x^3+ax^2+bx+c=0 ,
\eeq{1220}
and
\beq
3x^2+2ax+b=0,
\eeq{1221}
respectively, with
\beq
a=-M_{22},\qquad b={\L\tm_{11}^2\over 16}\pf m,\qquad
c=-\Big({\L\tm_{11}^2\over 32}\Big)^2 \det\l^2 .
\eeq{1222}
These are now the singularity equations of an elliptic curve
\beq
y^2=x^3+ax^2+bx+c .
\eeq{1223}
This curve defines the effective Abelian coupling in the Coulomb
branch.

We should note that when $m=0$ and $\tm_{11}\to 0$ (or $\det \l^2\to 0$),
the three singularities
of the elliptic curve degenerate; when $m\neq 0$ and $\tm_{11}\to 0$, two
out of the three singularities degenerate. This leads to a vacuum where
mutually non-local degrees of freedom are massless, similar to the
situation in pure $N=2$ supersymmetric $SU(3)$ gauge theories, considered
in \cite{ad}. Such a point might be interpreted as a non-Abelian Coulomb
phase \cite{S2}. (Integrating out the doublets, namely, taking $m\to\infty$
keeping $\L\pf m$ fixed, a similar phenomenon happens for the two vacua of
the $N_A=2$, $N_f=0$ theory: they collapse into a single vacuum where a
monopole and a dyon are mutually massless \cite{IS3}).

The reduction from $N_A=2$ to $N_A=1$ is obtained by the matching
\beq
\L_d^3={1\over 4}\L\tm_{11}^2 .
\eeq{1224}
Equation (\ref{1222}) becomes:
\beq
a=-M_{22}, \qquad b={\L_d^3\over 4} \pf m, \qquad c=-{\L_d^6\over 64}\det\l^2.
\eeq{1225}
This result is exactly the one obtained in the $N_A=N_f=1$ case
in eq. (\ref{1123}) (with $M_{22}$ replacing $M$, $\l^2$ replacing $\l$ and
$\L_d$ replacing $\L$).

Finally, we should remark that, unlike the $N_A=1$ cases,
eqs. (\ref{1215}), (\ref{1216}) are not the
singularity conditions of an elliptic curve, in general.
However, as mentioned before, they do become the singularity conditions of an
elliptic curve
when $\det\tm=0$ (as expected, physically, in a Coulomb phase), or when
$\det \l^1=0$ or when $\det\l^2=0$.\footnote{
Equations (\ref{1215}), (\ref{1216}) can be
reorganized in such a way that they become, manifestly, the singularity
conditions of an elliptic curve also when $\det \l^2=0$; equivalently, one
can use the $1\leftrightarrow 2$ symmetry to interchange 1 and 2 indices.}
Moreover, in eq. (\ref{1217}) we have identified a
physical meaning of the parameter $x$.

\noindent
\section{$SU(2)$ with $N_A=N_f=2$ ($b_1=0$) and other $b_1=0$ theories
revisited}
\setcounter{equation}{0}

So far, the (massless) cases with $b_1=0$ we have studied
($N_A=3, N_f=0$ in section 4, and $N_A=1, N_f=4$ in section 8) were
interacting conformal theories in the infra-red; the Yukawa couplings would
flow to values where the supersymmetry is enhanced
($N=4$, and $N=2$ with four flavors, respectively), and these theories have
an infra-red fixed line of marginal deformations. However, it is possible
that a $b_1=0$ theory flows to a free theory in an infra-red fixed point.
In this section, we consider this issue following ref. \cite{LS} and,
in particular, we argue that the $N_A=N_f=2$
case is a free theory in the infra-red.

We start by considering
an $N=1$ supersymmetric gauge theory with a simple gauge group,
$G$, and a gauge coupling, $g$, and with a tree-level superpotential
\beq
W_{tree}=\l\p_1\p_2\cdots\p_n ,
\eeq{221}
where the superfield $\p_i$ is in the representation $R_i$ of $G$.
The beta-functions are \cite{beta}
\beq
\b_{\l}\equiv {\partial\l(\m)\over \partial\ln\m}=
\l(\m)\Big(-3+\sum_{k=1}^nd(\p_k)+{1\over 2}\sum_{k=1}^n \g (\p_k)\Big)
\eeq{222}
\beq
\b_g\equiv {\partial g(\m)\over \partial\ln\m}=
-f(g(\m))\Big([3C(G)-\sum_{i}S(R_i)]+\sum_i S(R_i)\g (\p_i)\Big).
\eeq{223}
Here $f$ is a function of $g$, $d(\p_i)$ is the naive dimension
of $\p_i$, $\g(\p_i)$ is the anomalous dimension of $\p_i$, $C(G)$ is the
quadratic Casimir of the adjoint representation of $G$: $f^a_{cd}
f^{bcd}\equiv C(G)\d^{ab}$, and $S(R_i)$ is the Dynkin index of the
representation $R_i$: $\tr_{R}(T^aT^b)\equiv S(R)\d^{ab}$;
$3C(G)-\sum_{i}S(R_i)=b_1$.

Consider $G=SU(2)$ with $N_A\neq 0$ triplets and $N_f$ doublets
such that $b_1=0$, namely,
$(N_f,N_A)=(0,3),(2,2)$ or $(4,1)$, and with a superpotential of the
schematic form:
\beq
W_{tree}=\l\P QQ ,
\eeq{224}
where $\P$ is a triplet and $Q$ is a doublet (triplet) if $N_A\neq 3$
($N_A=3$).
(The superpotential includes several terms of this form if $N_A\neq 3$).
This is the massless case and, moreover, the operators in $W$ are marginal,
namely, $\sum_{k=1}^3 d(\p_k)=d(\P)+2d(Q)=3$. Therefore, for any Yukawa
coupling of the form (\ref{224}) the beta-function (\ref{222}) reads:
\beq
\b_{\l}={1\over 2}\l[\g(\P)+2\g(Q)] .
\eeq{225}
The gauge-coupling beta-function depends on the numbers $N_A$, $N_f$; we
consider the $b_1=0$ cases and, therefore, eq. (\ref{223}) reads:
\beq
\b_g=-f(g)[2N_A\g(\P)+N_f\g(Q)].
\eeq{226}

We now consider case by case:
\begin{itemize}
\item
$N_A=3$, $N_f=0$: $Q$ and $\P$ are triplets and, therefore, $\g(Q)=\g(\P)$,
which implies: $\b_g\sim\b_{\l}\sim\g(\P)$. Thus we get one equation in
two variables, so it has a fixed line of solutions: the space of
$N=4$, $SU(2)$ theories (with different gauge couplings).
\item
$N_A=1$, $N_f=4$: $\b_g\sim\b_{\l}\sim \g(\P)+2\g(Q)$. Therefore,
we get one equation in
two variables, so it has a fixed line of solutions: the space of
$N=2$, $SU(2)$ theories with four flavors (with different gauge couplings).
\item
$N_A=N_f=2$: $\b_g\sim 2\g(\P)+\g(Q)$, $\b_{\l}\sim \g(\P)+2\g(Q)$.
Therefore, we get two equations in two variables, and we expect
a discrete set of fixed points.
\end{itemize}

In all cases, $g=\l=0$ solves the equations and, for the
$N_A=N_f=2$ theory, no other solutions exists in a small enough
neighborhood  of this point. The sign of $\b_g$ is such
that it flows towards $g\to 0$ in the infra-red.
Therefore, we argue that the $N_A=N_f=2$, $SU(2)$ theory is infra-red
free,\footnote{
A related fact is that (unlike the $N_A=1$,
$N_f=4$ case) it is impossible to construct the matching
$``\L^{b_1}"=\a( \t_0 )f(\l^{\a})$ in a way that respects the global
symmetries.}
and there is not much more to say about it.

\noindent
\section{$SU(N_c)$ with $N_A=1$, $N_f=0$ ($b_1=2N_c$)}
\setcounter{equation}{0}
In the following sections we discuss $SU(N_c)$ ($N_c>2$) with $N_A$ matter
supermultiplets in the adjoint representation, and $N_f$ flavors ($N_f$
fundamentals and $N_f$ anti-fundamentals). We begin in this section
by integrating in a single adjoint matter ($N_A=1$) to pure $N=1$
supersymmetric  $SU(N_c)$ gauge theory.

The down theory has a nonperturbative superpotential (due to gluino
condensation):
\beq
W_d({\rm pure}\,\, N=1, SU(N_c))= N_c(\L_d^{b_{1,d}})^{1/N_c},
\eeq{n1}
where $b_{1,d}=3N_c$ is minus the one-loop coefficient of the gauge
coupling beta-function of the down theory. We now want to integrate in a
single supermultiplet in the adjoint representation, $\P^{ab}$,
$a,b=1,...,N_c$, $\tr \P=0$. The relevant gauge singlets, $U_k$,
are the $N_c-1$ Casimirs of $SU(N_c)$:
\beq
U_k=\tr \P^k , \qquad k=2,...,N_c ,
\eeq{n2}
and, therefore,
\beq
W_{tree}=\sum_{k=2}^{N_c} m_k U_k \implies W_{tree,d}\equiv
\sum_{k=2}^{N_c} m_k \tr \P^k |_{\langle \P\rangle}.
\eeq{n3}
Extremizing $W_{tree}$ with respect to $\P$, one should recall that $\P$ is
traceless and, therefore,
$\partial U_k/\partial \P^{ab}=k(\P^{k-1})^{ba}-(k/N_c)U_{k-1}\d_{ab}$.
{}From (\ref{n3}) we see that $W_{tree,d}$ is not unique, but a set of
solutions to polynomial equations in $U_k$, corresponding to different
classical vacua and, therefore, there could be several
branches. We will argue that a physical branch is found when
$W_{tree,d}=0$.\footnote{
It is possible that the other solutions also lead to
physical branches -- associated with other classical vacua, and maybe with
vacua such as those discussed in \cite{ad} -- whose $W_{tree,d}$ and $W_{\D}$
vanish when $m_k\to \infty$ for $k\neq 2$; it is plausible that such
branches involve these $m_k$, with $k>2$, also in the matching conditions,
in addition to $m_2$ \cite{efgr2}.}

Let us define the rescaled fields, $\teP$, with $(0,0)$ $U(1)_{\P}\times
U(1)_R$ quantum numbers
\beq
\teP\equiv {m_{N_c} \over m_{N_c-1}}\P,
\eeq{n4}
and the $N_c-2$ parameters, $t_k$, with $(0,0)$ $U(1)_{\P}\times
U(1)_R$ quantum numbers
\beq
t_1=\L {m_{N_c}\over m_{N_c-1}},
\qquad t_k=m_k{m_{N_c}^{N_c-1-k}\over
m_{N_c-1}^{N_c-k}}, \quad k=2,...,N_c-2 ,
\eeq{n5}
where $\L$ is the dynamically generated scale of the up theory.
We find that the $N_c-3$ parameters $t_2,...,t_{N_c-2}$ are involved in the
minimization of $W_{tree}$:
\beq
W_{tree,d}=\tau\Big[\sum_{k=2}^{N_c-2} t_k\tr \teP^k +
\tr\teP^{N_c-1}+\tr\teP^{N_c}\Big]_{\langle \teP \rangle}=\tau
f_{tree,d}(t_2,...,t_{N_c-2}),
\eeq{n6}
where the parameter $\tau$ has $(0,2)$ $U(1)_{\P}\times U(1)_R$
quantum numbers
\beq
\tau\equiv {m_{N_c-1}^{N_c}\over m_{N_c}^{N_c-1}} .
\eeq{n7}

The up theory has a nonperturbative superpotential
\ber
W_u(U_k)&=&[W_d+W_{tree,d}+W_{\D}-W_{tree}]_{\langle m_k\rangle} \nonumber\\
&=& \Big[(\L^{b_1})^{1/N_c} m_2
+\tau f(t) -\sum_{k=2}^{N_c} m_k U_k \Big]_{\langle m_k\rangle} ,
\eer{n8}
where $b_1=2N_c$,~\footnote{
In eq. (\ref{n8}) we wrote
$(\L^{2N_c})^{1/N_c}$ instead of $\L^2$, to keep the $N_c$ possibilities
corresponding to the $N_c$-roots of the identity:
$(\L^{2N_c})^{1/N_c}=\th_{N_c}^i\L^2$,
$i=0,1,...,N_c-1$, $\th_{N_c}\equiv \exp(2\pi i/ N_c)$.} and
\beq
t\equiv (t_1,...,t_{N_c-2}), \qquad
f(t)\equiv f_{tree,d}(t_2,...,t_{N_c-2})+f_{\D}(t) .
\eeq{n9}
In eq. (\ref{n8}) we  used the $U(1)_{\P}\times U(1)_R$ global symmetries to
write
\beq
W_{\D}=\tau f_{\D}(t) ,
\eeq{n10}
and we used the matching
\beq
\L_d^{b_{1,d}}=\Big({m_2\over N_c}\Big)^{N_c}\L^{b_{1}},
\eeq{n11}
where recall that $b_{1,d}=3N_c$ and $b_1\equiv b_{1,u}=2N_c$.

Unlike the $SU(2)$ case, when $N_c>2$ the limits $\L\to 0$ and $m_2\to
\infty$ are not enough to impose $W_{\D}=0$. However, it is shown in
the Appendix that
imposing in addition the condition to have a physical branch with a discrete
number of vacua implies that on such a branch $W_{tree,d}=0$, and $W_{\D}=0$.
This implies that  $W_u=0$ with the constraint:
\ber
U_2=(\L^{2N_c})^{1/N_c}\equiv \th_{N_c}^n\L^2,
& & n=0,1,..,N_c-1,\qquad \th_{N_c}=e^{{2\pi i\over N_c}},
\nonumber \\ U_k=0, & & k=3,...,N_c.
\eer{n12}
These correspond to the $N_c$ ``$SU(N_c)$ vacua'' which transform to each
other under a $Z_{N_c}$ transformation acting on the moduli space.

\noindent
\section{$SU(N_c)$ with $N_A=0$, $N_c>N_f\neq 0$ ($b_1=3N_c-N_f$)}
\setcounter{equation}{0}
The nonperturbative superpotential, $W_{N_f,0}$, of $N=1$ supersymmetric
$SU(N_c)$ gauge theory with $N_c>2$ and $N_f<N_c$ flavors
(when $N_f\geq N_c$ there are also baryons in the theory), can be
constructed \cite{study,S1}
just by the use of holomorphy and symmetries, or by integrating
in $N_f$ flavors to the pure $N=1$ supersymmetric $SU(N_c)$ gauge
theory with superpotential (\ref{n1}). This is done similarly to what we
described for $SU(2)$ in section 3; here we only present the result.

The $N_f$ flavors are $N_f$ matter supermultiplets
in the fundamental representation, $Q_i^a$, and $N_f$ supermultiplets in the
anti-fundamental, $\bQ_a^{\bi}$, $a=1,...,N_c$, $i,\bi=1,...,N_f$.
The relevant gauge singlets, $X_i^{\bi}$, are given in terms of $Q, \bQ$ by
\beq
X_i^{\bi}=Q_i^a \bQ_a^{\bi} .
\eeq{nf1}
The superpotential reads
\beq
W_{N_f,0}(X)=(N_c-N_f)\L^{3N_c-N_f\over N_c-N_f}(\det X)^{1\over N_f-N_c} +
\tr_{N_f} mX .
\eeq{nf2}

\noindent
\section{$SU(N_c)$ with $N_A=N_f=1$ ($b_1=2N_c-1$)}
\setcounter{equation}{0}
To derive the nonperturbative superpotential, $W_{1,1}$, of $N=1$
supersymmetric $SU(N_c)$ gauge theory with one supermultiplet in the
adjoint representation and one flavor, we integrate in an adjoint matter to
the supersymmetric $SU(N_c)$ theory with $N_f=1$.
The down theory superpotential is given by $W_{1,0}(X)$ in eq. (\ref{nf2}):
\beq
W_d=(b_1-N_c)\L^{b_1\over b_1-N_c}\Big({m_2\over N_c}\Big)^{N_c\over b_1-N_c}
X^{1\over N_c-b_1} .
\eeq{n111}
Here we used the matching (\ref{n11}) with $b_{1,d}=3N_c-1$ and $b_1\equiv
b_{1,u}=2N_c-1$.

The relevant gauge singlets we should add to $X=Q^a\bQ_a$ in the up theory
are $U_k$, given in eq. (\ref{n2}), and $Z$:~\footnote{
The number of microscopic degrees of freedom
(d.o.f($\P,Q,\bQ$) minus the gauge freedom) is $2N_c$, while the number of
macroscopic degrees of freedom (d.o.f($U_k,X,Z$)) is $N_c+1$. This means that
one might need to add the $N_c-1$ gauge singlets $Z_k\equiv Q\P^k\bQ$,
$k=2,...,N_c$ to the integrating in procedure. However, we checked that adding
$Z_k$ is irrelevant to the final result in the $SU(N_c)$ vacua branch
(see also the footnote after eq. (\ref{n1115})).}
\beq
Z=Q^a \P_a^b \bQ_b,
\eeq{n112}
where $\P$ is defined in section 11 and $Q,\bQ$ are defined
in section 12. Therefore,
\beq
W_{tree}=\sum_{k=2}^{N_c} m_k U_k + \l Z \implies
W_{tree,d}\equiv \Big[\sum_{k=2}^{N_c}m_k\tr\P^k+\l
Q\P\bQ\Big]_{\langle\P\rangle} ,
\eeq{n113}
namely, to find $W_{tree,d}$ we should solve the equation
\beq
{\partial W_{tree}\over \partial \P^t}=\sum_{k=2}^{N_c} km_k\P^{k-1}+\l
Q\bQ-{1\over N_c}\Big(\l X + \sum_{k=3}^{N_c} km_kU_{k-1} \Big) I=0 ,
\eeq{n114}
where $I$ is the $N_c\times N_c$ identity matrix.
The different solutions of eq. (\ref{n114}) correspond to
different branches of classical vacua of the theory.

We are interested in a
branch where $\P$ decouples as its mass approaches infinity: $m_2\to\infty$.
Therefore, in this limit, $W_i(m_2\to\infty)=W_{tree,d}+W_{\D}\to 0$.
Moreover, when $\L\to 0$, $W_{\D}(\L\to 0)\to 0$. Therefore, in the
combined limit $m_2\to\infty$ and $\L\to 0$ both $W_{\D}\to 0$ and
$W_{i}\to 0$, which implies that also $W_{tree,d}\to 0$. But $W_{tree,d}$
is independent of $\L$ and, therefore, we conclude that
$W_{tree,d}(m_2\to\infty)\to 0$. We refer to this branch as the
``perturbative branch.''

In the Appendix, it is shown that
requiring to have a branch with a discrete number of vacua, in addition to
an appropriate behavior in the $\L\to 0$ and $m_2\to\infty$ limits, is
consistent with a $W_{tree,d}$ evaluated at the single
$\langle\P\rangle$ solution of eq. (\ref{n114}) which is perturbative in
$\l/m_2$. This solution reads
\beq
\P={\l\over 2m_2}\Big({X\over N_c} I -
Q\bQ\Big)+{\cal O}\Big((\l/m_2)^2\Big) .
\eeq{n115}
Since
\beq
\tr_{N_c}(Q\bQ)^k=X^k ,
\eeq{n116}
the characteristic polynomial of the $N_c\times N_c$ matrix $Q\bQ$ is
\beq
(Q\bQ)^{N_c-1}(Q\bQ-X)=0 .
\eeq{n117}
This implies that $Q\bQ$ has an eigenvalue $X$ and $N_c-1$ zero
eigenvalues. Therefore, to solve (\ref{n114})
we can choose a pair of bases for which
\beq
Q\bQ=\diag (0,...,0,X) .
\eeq{n118}
In these bases, by using (\ref{n115}) one finds that
the perturbative solution to eq. (\ref{n114}) reads
\beq
\P=\diag (a,...,a,-(N_c-1)a) .
\eeq{n119}
This is the solution which corresponds classically to the $SU(N_c-1)$ vacua.
Using (\ref{n113}), (\ref{n114}) and (\ref{n119}) we find that in this
branch:
\beq
W_{tree,d}=-\sum_{k=2}^{N_c}(k-1)(N_c-1)\Big[1-(1-N_c)^{k-1}\Big]m_k \aa^k ,
\eeq{n1110}
where $\aa$ is the solution of
\beq
\l X-\sum_{k=2}^{N_c} k\Big[1-(1-N_c)^{k-1}\Big]m_k a^{k-1} = 0 ,
\eeq{n1111}
for which $\aa={\cal O}(\l X/m_2)$ as $m_2\to \infty$; there is a single
solution obeying this condition.

In the Appendix,  it is also shown that
requiring to have a branch with a discrete number of vacua, in addition to
an appropriate behavior in the $\L\to 0$ and $m_2\to\infty$ limits, implies
that such a physical branch has $W_{\D}=0$. Therefore, we find that in the
$SU(N_c)$ vacua branch, the nonperturbative superpotential of the up theory
is derived by
\beq
W_u=[W_d+W_{tree,d}-W_{tree}]_{\langle m_k\rangle,\langle\l\rangle} ,
\eeq{n1112}
where $W_d$, $W_{tree,d}$ and $W_{tree}$ are given in eqs. (\ref{n111}),
(\ref{n1110}) and (\ref{n113}), respectively. After some algebra one finds
\beq
W_{1,1}(U_2,X,Z)=-\L^{-b_1}X\G^{N_c}+\sum_{k=2}^{N_c} m_kU_k+mX+\l Z ,
\eeq{n1113}
where recall $b_1=2N_c-1$, and
\beq
\G=U_2-{N_c\over N_c-1} x^2, \qquad x=ZX^{-1} ,
\eeq{n1114}
and the $N_c-2$ constraints:
\beq
U_k={(1-N_c)^{k-1}-1\over (1-N_c)^{k-1}} x^k, \qquad k=3,...,N_c .
\eeq{n1115}
Equation (\ref{n1115}) is a set of classical conditions (to check it,
on the $SU(N_c-1)$ classical vacua,
one may use eqs. (\ref{n118}), (\ref{n119})), while $\G$ in eq.
(\ref{n1114}) vanishes classically, as expected physically due to the
negative power of $\L$ in the superpotential (\ref{n1113}).
If one would ignore the ``apparently
irrelevant'' operators $U_k$ with $k>2$, in the integrating in procedure, one
would fail to get $\G$ which vanishes classically. In other words, one gets
for $\G$ the characteristic polynomial for $\P$ (with $\P$ being replaced by
$x$, up to an overall $x$-dependent factor); equation (\ref{n1114}) is
the value of the characteristic polynomial on the classical constraints in
eq. (\ref{n1115}), while ignoring the $U_k$ with $k>2$ means to set their
values to zero in the characteristic polynomial, thus leading to an object
that does not vanish classically.\footnote{
Unlike the $U_k$, the $Z_k$ with $k>2$, defined in the footnote before eq.
(\ref{n112}),  are indeed {\em irrelevant}.
We checked that adding them to the integrating in procedure does not affect
the result that $W_{\D}=0$ and, consequently, does not change the final
result in eqs. (\ref{n1113})-(\ref{n1115}), but give extra (classical)
constraints for $Z_k$: $Z_kX^{-1}=x^k$.}

We now want to find the vacua of the theory in the branch (\ref{n1113}),
(\ref{n1115}), namely, we should solve the equations of motion
$\d W_{1,1}/\d U_k=\d W_{1,1}/\d X = \d W_{1,1}/\d Z=0$
on the constraints (\ref{n1115}). We study here only the case
\beq
m_k=0,\quad k=3,...,N_c .
\eeq{n1116'}
The equations of motion read\footnote{In the presence of tree-level terms
with $m_k\neq 0$ for $k\geq 3$, the equations of motion receive
$m_k$-dependent corrections, due to the constraints (\ref{n1115}), and one
gets a different vacua structure \cite{efgr2}: turning on
$\sum_{k=3}^{N_c}m_kU_k$ give extra vacua not considered here.}:
\beq
m_2=N_c\L^{-b_1}X\G^{N_c-1} ,
\eeq{n1116}
\beq
m=\L^{-b_1}\G^{N_c-1}\Big(\G-N_c x {\partial\G\over \partial x}\Big) ,
\eeq{n1117}
\beq
\l=N_c\L^{-b_1}\G^{N_c-1}{\partial\G\over \partial x} .
\eeq{n1118}
Combining eq. (\ref{n1117}) with eq. (\ref{n1118}) we get
\beq
\G^{N_c}=\L^{b_1}\l(x+\m) , \qquad \m=\l^{-1} m .
\eeq{n1119}
Equations (\ref{n1119}) and (\ref{n1118}) are the singularity conditions of
a genus $N_c-1$ hyperelliptic curve defined by
\beq
y^2=\G(x)^{N_c}-\L^{b_1}\l(x+\m) .
\eeq{n1120}

Using eqs. (\ref{n1118}), (\ref{n1119}) to solve $U_2$ in terms of $x$,
and eliminating $U_2$ in eq. (\ref{n1118}) we find
\beq
x^{N_c}(x+\m)^{N_c-1}-\Big({1-N_c\over 2N_c^2 }\Big)^{N_c} \L^{b_1}\l=0 ,
\eeq{n1121}
and
\beq
U_2=-{N_c\over N_c-1}x(b_1x+2N_c\m) .
\eeq{n1122}
Therefore, we find that $W_{1,1}$ (\ref{n1113}) has
$b_1=2N_c-1=N_c+(N_c-1)N_f$ vacua,
namely, the $2N_c-1$ solutions for $M(x)$ in terms of the $2N_c-1$
roots of the polynomial equation for $x$ (\ref{n1121}) -- the
singularities of the hyperelliptic curve (\ref{n1121}) --
and the solution for $X$ given by eq. (\ref{n1116}) ($Z$ is now determined
by $Z=xX$, and recall that $U_k, k>2$ are fixed by (\ref{n1115})).
These $b_1=2N_c-1$ vacua are the vacua of the theory in the
Higgs/confinement branch\footnote{
Adding a tree-level superpotential $\sum_{k=3}^{N_c} m_k U_k$ gives rise,
generically, to a total of $N_c^2-1$ solutions \cite{efgr2}; the extra
$N_c(N_c-2)$ solutions go to infinity in the $U_k$ space when $m_k/m_2\to
0$, $k\geq 3$, and one is left with the $2N_c-1$ ``$SU(N_c)$ vacua''
considered here. Moreover, adding a tree-level superpotential
$\sum_{k=2}^{N_c}\l_k Z_k$ gives rise, generically, to a total of
$2N_c(N_c-1)$ solutions \cite{efgr2}; this is due to the constraints
discussed in a footnote after eq. (\ref{n1115}).}.
The phase transition points to the Coulomb branch
are at $X=0$. This happens iff the adjoint superfield is massless, namely
\beq
X=0 \Leftrightarrow m_2=0.
\eeq{n1123}
The values of $U_k$ at the $SU(N_c)$ vacua are independent of the value
$X$. When $m=m_k=0$, there is a $\Z_{2N_c-1}$ transformation relating the
different vacua; this is a symmetry of the $U_k$ moduli space in the
Coulomb phase.

\noindent
\section{$SU(2)$ with $N_A=N_f=1$ revisited}
\setcounter{equation}{0}

In this section, we rederive the results of section 5 in a simpler way,
similar to the manipulation for $N_c>2$ in section 13.

For $N_c=2$, the anti-fundamental representation is equivalent to the
fundamental representation and, therefore, $Z$ of eq. (\ref{n112}) becomes
a symmetric $2\times 2$ matrix (see eq. (\ref{XMZ})).
Moreover, $X$ of section 13 denotes $\pf X$ where $X$ is an
antisymmetric $2\times 2$ matrix, and
there is a single Casimir which  we denote by $U_2\equiv M$,
as in section 5. We thus find that the superpotential $W_{1,1}$ is given
by eq. (\ref{n1113}) with (\ref{n1114}) replaced by
\beq
\G=M-2x^2=M+\tr(ZX^{-1})^2, \qquad  x=(\det Z)^{1/2} (\pf X)^{-1} .
\eeq{r1}
Following the discussion in section 13, we find that the vacua are
given by the solutions to  eqs. (\ref{n1116}), (\ref{n1117}), and eq.
(\ref{n1118}) is modified to
\beq
(\det\l)^{1/2}=\L^{-3}\G{\partial\G\over \partial x} .
\eeq{r2}
Namely, we find that $X$ is solved by
\beq
\pf X={m_2\L^3\over 2\G}
\eeq{r3}
($m_2\equiv \tm$ in the notations of section 5),
and $x$, $M$ are given by the singularity conditions of an elliptic curve:
\beq
y^2=\G^2-\L^3 (\a x+m) , \qquad \a\equiv 2(\det\l)^{1/2}.
\eeq{r4}
The curve in the form (\ref{r4}) was presented in ref. \cite{sun}.
This elliptic curve is related to the previous one, in eqs.
(\ref{1122}), (\ref{1123}), by rescaling $m\to m/2$, $\l\to \l/\sqrt{2}$ ,
together with an $SL(2,\C)$ transformation:
\beq
x\to {ax+b\over cx+d}, \qquad y\to  K(cx+d)^2 y, \qquad ad-bc=1,
\eeq{r5}
where $a,b,c,d$ and $K$ are given in terms of $M, \a\L^3$ and $m\L^3$.

\noindent
\section{$SU(N_c)$ with $N_A$, $N_f<N_c$ ($b_1=3N_c-N_cN_A-N_f$)}
\setcounter{equation}{0}

In this section, we  present the effective superpotential,
$W_{N_f,N_A}$, in $N=1$ supersymmetric $SU(N_c)$ gauge theory, $N_c>2$,
with $N_A$ matter superfields in the adjoint representation,
$\P_{\a}^{ab}$, $\tr\P=0$, and $N_f<N_c$ flavors, $Q_i^a$, $\bQ_a^{\bi}$
(when $N_f\geq N_c$ there are also baryons in the theory). Here
$a,b=1,...,N_c$, $i,\bi=1,...,N_f$, and $\a=1,...,N_c$.
As before, we derive the superpotential by integrating in
$N_A$ adjoint supermultiplets to a supersymmetric $SU(N_c)$ gauge theory
with $N_f<N_c$ flavors, presented in section 12; the superpotential of
the down theory is $W_d=W_{N_f,0}(X)$, given in eq. (\ref{nf2}).
We consider the up theories
with one-loop asymptotic freedom or conformal invariance, for which
\beq
b_1=3N_c-N_f-N_cN_A\geq 0 ,
\eeq{f1}
where $-b_1$ is the one-loop coefficient of the gauge coupling beta-function.

The relevant gauge singlets we should add to $X_i^{\bi}$ in eq. (\ref{nf1})
are
\ber
U_{(\a_1,..,\a_k)}&=&\tr_{N_c}(\P_{\a_1}\cdots \P_{\a_k}), \qquad
k=2,...,N_c,\quad \a_n=1,...,N_A , \nonumber \\
Z_{i\,\a}^{\bi} &=& \tr_{N_c} (Q_i\bQ^{\bi}\P_{\a}) .
\eer{f2}
(For $N_A=1$, the gauge singlets $Z_{i\,(\a_1,..,\a_k)}^{\bi}=\tr_{N_c}
(Q_i\bQ^{\bi}\P_{\a_1}\cdots\P_{\a_k})$, $k=2,...,N_c$,
are irrelevant, as in the $N_f=1$ case; they do not change the final result
even if added to the integrating in procedure\footnote{
We did not discuss here the baryon-like operators of refs. \cite{K,KS};
we checked that the operators, containing at most $N_c$ adjoint superfields,
are irrelevant for the integrating in procedure on the perturbative
branch (although they might be important on the nonperturbative branches
\cite{efgr2}).}.
This is assumed also when $N_A=2$.)
We obtain the superpotential\footnote{When $b_1=2N_c$, the nonperturbative
superpotential vanishes and one obtains an additional constraint; this
happens only in case $N_A=1$, $N_f=0$, considered in section 11.}
\ber
W_{N_f,N_A}(X,U,Z)&=&(b_1-2N_c)\Big[\L^{-b_1} {\rm det}_{N_f} X
({\rm det}_{N_A} \G)^{N_c}\Big]^{1\over 2N_c-b_1} \nonumber \\
&+& \sum_{k=2}^{N_c} \Big(\sum_{\a_1=1}^{N_A}\dots \sum_{\a_k=1}^{N_A}
m_{(\a_1,..,\a_k)}U_{(\a_1,..,\a_k)}\Big) \nonumber\\
&+& \tr_{N_f} mX + \tr_{N_f} \l^{\a} Z_{\a} ,
\eer{f3}
and the constraints
\ber
U_{(\a_1,..,\a_k)}&=&\tr_{N_f}(Z_{\a_1}X^{-1}\cdots Z_{\a_k}X^{-1})
\nonumber\\
&-& {1\over (N_f-N_c)^{k-1}} \tr_{N_f}(Z_{\a_1}X^{-1})\cdots
\tr_{N_f}(Z_{\a_k}X^{-1}) ,
\eer{f4}
where
\beq
\G_{\a\b}=U_{(\a,\b)}-\tr_{N_f}(Z_{\a}X^{-1}Z_{\b}X^{-1})-{1\over
N_c-N_f}\tr_{N_f} (Z_{\a}X^{-1}) \tr_{N_f} (Z_{\b}X^{-1}) .
\eeq{f5}

To get this result, for $N_A=1$, we follow the strategy used in the
$N_A=N_f=1$ case in  section 13. Namely, we use limiting considerations and
impose the physical condition to have a finite number of $SU(N_c)$ vacua
branches, to find that $W_{\D}=0$. Then, using the perturbative branch, where
$W_{tree,d}(m_{(\a,\b)}\to\infty)\to 0$, and after some algebra, we find
eqs. (\ref{f3})-(\ref{f5}) for $N_A=1$.

The $N_A=3$, $N_f=0$ case includes the $N=4$ supersymmetric $SU(N_c)$ gauge
theory. As for the $SU(2)$ case with $N_A=3$, discussed in section 4,
in this case, $W$ must be equal to the Yukawa coupling tree-level
term of the three adjoint superfields. This is, indeed, the result in eq.
(\ref{f3}) for $N_A=3$, $N_f=0$.
Moreover, by integrating out a single adjoint superfield, one
obtains the result in (\ref{f3}) for $N_A=2$, $N_f=0$.
The superpotential for $N_A=2$, $N_f\neq 0$ in eq. (\ref{f3}) is
conjectured. To get this result, we use the assumption that $W_{\D}=0$ in the
perturbative branch also in this case.

One may now find the quantum vacua of the theory, by solving the equations
of motion derived from (\ref{f3}).
In case there is a single adjoint matter superfield ($N_A=1$, $N_f<N_c$),
and setting
\beq
m_{(\a_1,..,\a_k)}=0 , \qquad k>2 ,
\eeq{f6'}
we find that the number of (branches of) $SU(N_c)$ vacua is
\ber
{\rm no.}\,\,{\rm of}\,\,  SU(N_c)\,\, {\rm vacua} & & {\rm for}\,\,
N_f<N_c:\nonumber\\
{1\over 2}b_1(N_f+1)&=&N_c+N_f(N_c-1)-{1\over 2}N_f(N_f-1) .
\eer{f6}
When, in addition,  $m=m_{(\a_1,\a_2)}=0$, there is a $\Z_{b_1}$ symmetry
relating the different vacua.  This symmetry
can be read directly from the quantum superpotential (\ref{f3}):
it acts on $Q, \bQ$ and $\P$ by
\beq
\P\to e^{{2\pi i n\over b_1}}\P, \qquad Q\to e^{-{\pi i n\over b_1}}Q, \quad
\bQ\to e^{-{\pi i n\over b_1}}\bQ, \qquad n=1,...,b_1 ,
\eeq{f7}
and leave invariant both the tree-level term, $\tr_{N_f}\l^{\a}Z_{\a}$,
and the nonperturbative superpotential,
$(\L^{-b_1} {\rm det}_{N_f} X ({\rm det}_{N_A} \G)^{N_c})^{1/N_f}$.

What about the duality of refs. \cite{K,KS}?
It is valid when $N_c>2$ and $N_f\geq N_c/(k-1)$, $k=3,...,N_c$, depending
on which $\tr\P^k$ interaction is turned on, and at the infra-red fixed
point of the renormalization group flow.
Yet, in the $SU(N_c)$ vacua branches, considered here, we do not
have the tree-level couplings, $m_{(\a_1,..,\a_k)}$ with $k>2$, which
are required for the duality arguments of \cite{K,KS}. Studying the cases
where $m_{(\a_1,..,\a_k)}\neq 0$, as well as other branches
of $SU(N_c)$ supersymmetric gauge theories, might be useful to
understand this duality \cite{efgr2}.

\vskip .3in \noindent
{\bf Acknowledgements} \vskip .2in \noindent
We thank N. Seiberg for discussions.
The work of SE is supported in part by the BRF - the Basic Research
Foundation. The work of AG is supported in part by BSF - American-Israel
Bi-National Science Foundation, by the BRF,  and by an Alon fellowship.
The work of ER is supported in part by BSF and by the BRF.

%\newpage
\vskip .3in

\section*{Appendix A - $W_{\D}=0$}
\setcounter{equation}{0}
\renewcommand{\theequation}{A.\arabic{equation}}

In this Appendix, we show in detail the considerations leading to the
conclusion that $W_{\D}=0$ in the examples considered in
sections 3,4,11,13.

\subsection*{A.1 -
Down theory = $SU(2)$ with $N_A=0$, $N_f\leq 4$, Up theory
= $SU(2)$ with $N_A\neq 0$}

The $U(1)_Q\times U(1)_{\P}\times U(1)_R$ quantum numbers of $W$ are
$(0,0,2)$ and, therefore,
\beq
W_{\D}(X,\L,\tm,\l)=W_{tree,d}f(t),
\eeq{a1}
where $W_{tree,d}$ is given in eq. (\ref{Wtdfa}), and $t$ denotes,
schematically, any singlet of the $SU(2N_f)$ flavor symmetry with $(0,0,0)$
$(Q,\P,R)$-charges.
When integrating in triplets to an $SU(2)$ theory with doublets,
$W_{\D}$ depends on $X, \l, \L$ and $\tm$.
The quantum numbers of $X, \l,\L^{b_1}, \tm$ are
\ber
X&:& (Q,\P,R)=(2,0,0), \nonumber\\
\l&:& (Q,\P,R)=(-2,-1,2), \nonumber\\
\L^{b_1}&:& (Q,\P,R)=(2N_f,4N_A,4-4N_A-2N_f), \qquad b_1=6-2N_A-N_f,
\nonumber\\
\tm&:& (Q,\P,R)=(0,-2,2) .
\eer{a2}
Therefore, if we denote schematically,
\beq
t\sim (\L^{b_1})^a\tm^b X^c \l^d ,
\eeq{a3}
we find that the condition that $t$ has $(0,0,0)$ $(Q,\P,R)$-charges implies
\beq
(2+2N_A-N_f)a=b .
\eeq{a4}

We now want to impose the limits:
\beq
W_{\D}(\tm\to\infty)\to 0, \qquad W_{\D}(\L\to 0)\to 0 .
\eeq{a5}
Therefore, we are only interested in the dependence of $W_{\D}$
on $\tm$ and $\L$. Recall that, by definition, $W_{tree,d}$ is
$\L$-independent. Using the schematic dependence of $W_{tree,d}$
(\ref{Wtdfa}) on $\tm$
\beq
W_{tree,d}(\tm)\sim {1\over \tm},
\eeq{a6}
and analyzing the several cases with $N_f\neq 0$, $b_1\geq 0$ we find the
following schematic $\tm,\L$ dependence:
\begin{itemize}
\item
For $N_A=1$, $N_f=1,2,3$ ($b_1=4-N_f$):
equation (\ref{a4}) implies $b=b_1a$ and, therefore,
\beq
W_{\D}(\tm,\L)\sim {1\over \tm} f\left((\tm \L)^{b_1}\right) .
\eeq{a7}
\item
For $N_A=1$, $N_f=4$ ($b_1=0$): $``\L^{b_1}" \sim (\det\l)^{-1/2}$ (see
eq. (\ref{4111'})) and, therefore, $t\sim \tm^bX^c\l^d$. The condition
that $t$ has $(0,0,0)$ $(Q,\P,R)$-charges implies $b=c=d=0$ and,
therefore, $t$ can only depend on $\tau_0$ (introduced in section 8).
We thus find
\beq
W_{(0,0,2)}\sim {(\l X)^2\over \tm}f(\tau_0)\sim {X^2\over \L_d} f(\tau_0) ,
\eeq{a8'}
where $\L_d\equiv \L_{N_f=4,N_A=0}\sim \tm (\det\l)^{-1/4}$ (see section 8).
Using (\ref{a8'}), holomorphy and $SU(8)$ flavor symmetry we find that
\beq
W_{\D}\sim W_{tree,d}f(\tau_0) .
\eeq{a8}
The function $f(\tau_0)$ is related to the function $\b(\tau_0)$ used in
section 8 to rescale $M$; here we can get rid of $f$ in $W_{\D}$ by
rescaling $\tm\to (1+f)\tm$ together with $M\to M/(1+f)$, which takes
$W_{tree,d}+W_{\D}\to W_{tree,d}$ while leaving $\tm M$ invariant.
\item
For $N_A=2$, $N_f=1$ ($b_1=1$):
equation (\ref{a4}) implies $b=5a$ and, therefore,
\beq
W_{\D}\sim {1\over \tm} f(\tm^5 \L),
\eeq{a9}
\item
For $N_A=N_f=2$ ($b_1=0$): the theory is infra-red free (see section 10).
\end{itemize}
Since we can trust the instanton expansion in the Higgs branch, we know that
$W_{\D}$ depends on integer powers of $\L^{b_1}$. Therefore, the limits
(\ref{a5}) and eqs. (\ref{a7}), (\ref{a8}), (\ref{a9}) imply that for all
infra-red nontrivial cases with $N_f\neq 0$, the intermediate
superpotential $W_i$ (\ref{WD})  behaves like $W_{tree,d}$. As we
already included $W_{tree,d}$ in the procedure, we conclude that $W_{\D}=0$
when integrating in $N_A$ triplets.
Finally, when $N_f=0$, it is easy to show that $W_{\D}=0$.

\subsection*{A.2 -
Down theory = $SU(N_c)$ with $N_A=N_f=0$, Up theory =
$SU(N_c)$ with $N_A=1$}

{}From eq. (\ref{n8}) we obtain:
\ber
U_2-(\L^{b_1})^{1/N_c} &=& \O^2\partial_2 f , \nonumber\\
U_k &=& \O^k \partial_k f, \qquad k=3,...,N_c-2, \nonumber\\
U_{N_c-1} &=& \O^{N_c-1}\Big[N_c f-t_1\partial_1 f-\sum_{k=2}^{N_c-2}
(N_c-k)t_k\partial_k f\Big],\nonumber\\
U_{N_c} &=& \O^{N_c}\Big[-(N_c-1)f+t_1\partial_1 f+\sum_{k=2}^{N_c-2}
(N_c-k-1)t_k\partial_k f\Big],\nonumber\\
\eer{1a1}
where
\beq
\O\equiv {\L \over t_1}, \qquad \partial_k f\equiv {\partial f\over
\partial t_k}
\eeq{1a2}
($f=f_{tree,d}+f_{\D}$ is given in eq. (\ref{n9})).
On the solution (\ref{1a1}) we find that $W_u=0$ for any $f$. The $N_c-1$
equations for $U_k$, $k=2,...,N_c$, in terms of the $N_c-2$ variables $t$
(see eqs. (\ref{n9}), (\ref{n5})) define, in general, an $N_c-2$
dimensional manifold of vacua. There will be a discrete set of vacua only
if the $U_k$ in eq. (\ref{1a1}) turn out to be $t$-independent. This
happens only for $f$ which solves the equations:
\ber
\partial_k f=C_k t_1^k, \qquad k=2,...,N_c-2, \nonumber\\
N_c f-t_1\partial_1 f-\sum_{k=2}^{N_c-2} (N_c-k)t_k\partial_k
f=C_{N_c-1}t_1^{N_c-1},\nonumber\\
-(N_c-1)f+t_1\partial_1 f+\sum_{k=2}^{N_c-2}
(N_c-k-1)t_k\partial_k f=C_{N_c}t_1^{N_c} ,
\eer{1a3}
where $C_k$ are independent of $t$ and $\L$ (the $\L$-independence follows
from $U(1)_{\P}\times U(1)_R$ charge conservation). For $N_c>3$ ($N_c=3$ will
be considered separately), the general solution is
\beq
f=\sum_{k=2}^{N_c-2} C_k t_1^k t_k,  \qquad C_{N_c-1}=C_{N_c}=0 .
\eeq{1a4}
In the limit $\L\to 0$, the parameter $t_1$ defined in (\ref{n5}) goes to
zero and, therefore, $f\to 0$. Since we impose $W_{\D}(\L\to 0)\to
0$, we find $f_{\D}(\L\to 0)\to 0$ and, therefore, $f_{tree,d}(\L\to 0)\to
0$. But $f_{tree,d}(t_2,...,t_{N_c-2})$ is independent of $\L$ (because
$t_2,...,t_{N_c}$, defined in (\ref{n5}), are $\L$-independent) and,
therefore
\beq
f_{tree,d}=0 .
\eeq{1a5}
Equation (\ref{1a5}) means that the condition to have a discrete set of
vacua chooses the branch where $W_{tree,d}=0$.\footnote{
It can be verified that this is the only branch of $W_{tree,d}$
which obeys: $W_{tree,d}(m_2\to\infty)\to 0$.}

For $N_c>3$, the parameter $\tau$, defined in (\ref{n7}), is $m_2$
independent. Therefore, eq. (\ref{n10}) and the condition
$W_{\D}(m_2\to\infty)\to 0$ imply $f_{\D}(m_2\to\infty)\to 0$. In the limit
$m_2\to\infty$, we get from (\ref{n5}) that $t_2\to\infty$ while
$t_1,t_3,...,t_{N_c-2}$ are anything. Therefore, we conclude that
$C_2=...=C_{N_c-2}=0$, which implies
\beq
f_{\D}=0 .
\eeq{1a6}
We thus found that, for $N_c>3$, $W_{\D}=W_{tree,d}=0$, and from eq.
(\ref{1a1}) with $f=0$ one finds eq. (\ref{n12}).

Finally, for $N_c=3$, eqs. (\ref{n7}), (\ref{n10}) and (\ref{1a3}) imply
\beq
f(t_1)=C_2 t_1^2+C_3 t_1^3 .
\eeq{1a7}
As before, $W_{\D}(\L\to 0)\to 0$ implies $W_{tree,d}=0$ and, therefore,
$W_{\D}=(C_2t_1^2+C_3t_1^3)m_2^3/m_3^2$. Now, the limit
$W_{\D}(m_2\to\infty)\to 0$ implies $C_2=C_3=0$ and, therefore, $W_{\D}=0$.

\subsection*{A.3 -
Down theory = $SU(N_c)$ with $N_A=0$, $N_f=1$, Up theory =
$SU(N_c)$ with $N_A=N_f=1$}

In the integrating in procedure (\ref{n1112}) we involve $N_c+2$
parameters: $X, \L, \l$ and $m_k$, $k=2,...,N_c$. Therefore, we can
construct $N_c-1$ parameters with $(0,0,0)$ $U(1)_Q\times U(1)_{\P}\times
U(1)_R$ quantum numbers:
\ber
t_0=\l\L^{b_1}\Big({m_{N_c}\over m_{N_c-1}}\Big)^{b_1}, \, & &\,
t_1=\l X {m_{N_c}^{N_c-2}\over m_{N_c-1}^{N_c-1}}, \nonumber\\
t_k=m_k{m_{N_c}^{N_c-1-k}\over m_{N_c-1}^{N_c-k}}, \, & &\, k=2,...,N_c-2 .
\eer{2a1}
We also define the parameter $\tau$ with $(0,0,2)$ $(Q,\P,R)$-charges
\beq
\tau\equiv {m_{N_c-1}^{N_c}\over m_{N_c}^{N_c-1}} .
\eeq{2a2}
Let us denote
\beq
f(t)\equiv f_d(t_0,t_1,t_2)+f_{tree,d}(t_1,...,t_{N_c-2})+f_{\D}(t), \qquad
t\equiv (t_0,...,t_{N_c-2}),
\eeq{2a3}
where $f_d,f_{tree,d}$ and $f_{\D}$ are defined by
\beq
W_d=\tau f_d( t_0,t_1,t_2), \qquad
W_{tree,d}=\tau f_{tree,d}(t_1,...,t_{N_c-2}), \qquad
W_{\D}=\tau f_{\D}(t) .
\eeq{2a4}
{}From eqs. (\ref{n111}), (\ref{2a1}) we read:
\beq
f_d(t_0,t_1,t_2)=(N_c-1)N_c^{{-N_c\over N_c-1}}\Big({t_0 t_2^{N_c}\over
t_1}\Big)^{1\over N_c-1} .
\eeq{2a5}
The function $f_{tree,d}$ can be a' priori any branch of $W_{tree,d}$ in
(\ref{n113}), (\ref{n114}).
Since $W_{\D}(\L\to 0)\to 0$, and as we trust the instanton
expansion in the Higgs phase, $f_{\D}$ is holomorphic in $t_0$ and
$f_{\D}(t_0\to 0)\to 0$. Therefore,
\beq
f_{\D}(t)=\sum_{n=1}^{\infty} a_n(t_1,...,t_{N_c-2})t_0^n ,
\eeq{2a6}
and, moreover, for $N_c>3$ ($N_c=3$ will be considered separately)
\beq
W_{\D}(m_2\to\infty)\to 0 \Leftrightarrow f_{\D}(t_2\to\infty)\to 0
\Leftrightarrow a_n(t_2\to\infty)\to 0 .
\eeq{2a7}

Now, the integrating in procedure (\ref{n1112}) reads:
\beq
W_u=\Big[\tau f-\sum_{k=2}^{N_c} m_k U_k -\l Z\Big]_{\langle
m_k\rangle,\langle\l\rangle} ,
\eeq{2a8}
and we obtain
\ber
U_k&=&\O^k\pa_k f, \qquad k=2,...,N_c-2 ,\nonumber\\
U_{N_c-1}&=&\O^{N_c-1}\Big[N_c f-b_1 t_0 \pa_0 f
-\sum_{k=1}^{N_c-2}(N_c-k)t_k\pa_k f\Big], \nonumber\\
U_{N_c}&=&\O^{N_c}\Big[-(N_c-1)f+b_1 t_0 \pa_0 f
+\sum_{k=1}^{N_c-2}(N_c-1-k)t_k\pa_k f\Big], \nonumber\\
x&=&{\O\over t_1}[t_0\pa_0 f+t_1\pa_1 f] ,  \qquad x=ZX^{-1}
\eer{2a9}
where
\beq
\O\equiv \Big({\l\L^{b_1}\over t_0}\Big)^{1/b_1} , \qquad \pa_k f\equiv
{\pa f\over \pa t_k}.
\eeq{2a10}
Using (\ref{2a10}) and the definitions, we get
\beq
W_u=\tau(-t_0\pa_0 f)|_{\langle m_k\rangle,\langle\l\rangle}
=-\L^{-b_1}Xx^{2N_c}B_1(t)|_{\langle m_k\rangle,\langle\l\rangle} ,
\eeq{2a11}
where
\beq
B_1(t)={t_0^2 t_1^{b_1}\pa_0 f\over (t_0\pa_0 f+t_1\pa_1 f)^{2N_c}} .
\eeq{2a12}

Eliminating $\l$ from eq. (\ref{2a9}) we get
\beq
{U_k\over x^k}=B_k(t), \qquad k=2,...,N_c ,
\eeq{2a13}
where
\beq
B_k(t)={t_1^k\pa_k f\over (t_0\pa_0 f+t_1\pa_1 f)^k} , \qquad
k=2,...,N_c-2 ,
\eeq{2a14}
\beq
B_{N_c-1}(t)={t_1^{N_c-1}\Big[N_c f-b_1 t_0 \pa_0 f
-\sum_{k=1}^{N_c-2}(N_c-k)t_k\pa_k f\Big]\over (t_0\pa_0 f+t_1\pa_1
f)^{N_c-1}} ,
\eeq{2a15}
\beq
B_{N_c}(t)={t_1^{N_c} \Big[-(N_c-1)f+b_1 t_0 \pa_0 f
+\sum_{k=1}^{N_c-2}(N_c-1-k)t_k\pa_k f\Big]\over  (t_0\pa_0 f+t_1\pa_1
f)^{N_c}} .
\eeq{2a16}
Equations (\ref{2a14}), (\ref{2a15}), (\ref{2a16}) are $N_c-1$ equations
with $N_c-1$ parameters $t\equiv (t_0,...,t_{N_c-2})$. So, in principle, we
can solve $t$ in terms of $B_k$, $k=2,...,N_c$, and insert in (\ref{2a12}),
(\ref{2a11}) to get $W_u$ in terms of $B_k$, $k=2,...,N_c$.

The equations of motion for $W=W_u+mX+\l Z$ with respect to variation of
$X$ and $Z$ lead (after taking their combinations) to eqs. (\ref{n1119}),
(\ref{n1118}) (with $\G^{N_c}$ being replaced by $x^{2N_c}B_1$), namely,
\ber
x^{2N_c}B_1=\L^{b_1}(m+\l x) , \nonumber\\
{\pa \over \pa x}( x^{2N_c}B_1)=\L^{b_1}\l .
\eer{2a17}
Since $B_1$ depends on $B_k$ with $k\geq 2$, these equations define,
generically,  a surface in the $B_k$ space. This ``surface'' is a discrete
set of points (vacua) iff $B_1$ depends on one $B_k$. Without loss of
generality, we can express all $B_k$'s in terms of $B_2$:
\beq
{\rm discrete}\,\, {\rm set}\,\, {\rm of}\,\, {\rm vacua}\, \Leftrightarrow
B_1=F_1(B_2), \quad B_k=F_k(B_2), \quad k=3,...,N_c .
\eeq{2a18}

We can now use the properties of the $\L\to 0$ and $m_2\to \infty$ limits
to show that a discrete set of vacua is obtained iff when
$W_{tree,d}$ is in the $\l/m_2$ perturbative branch (see section 13)
then $W_{\D}=0$.

In the limit $\L\to 0$, {\em i.e.}, $t_0\to 0$ (recall eq. (\ref{2a1})),
we impose $W_{\D}(\L\to
0)\to 0$ and, therefore, $f_{\D}(t_0\to 0)\to 0$. Moreover, (\ref{2a5})
implies that $f_d(t_0\to 0)\to 0$. We denote
\beq
B_k^{(0)}\equiv B_k(t_0=0,t_1,...,t_{N_c-2}) ,
\eeq{2a19}
and for $t_0=0$ eqs. (\ref{2a12}), (\ref{2a14}), (\ref{2a15}) and
(\ref{2a16}) read:
\beq
B_1^{(0)}=0 ,
\eeq{2a20}
\beq
B_k^{(0)}={\pa_k f_{tree,d}\over (\pa_1 f_{tree,d})^k}, \qquad
k=2,...,N_c-2 ,
\eeq{2a21}
\beq
B_{N_c-1}^{(0)}={N_c f_{tree,d}-\sum_{k=1}^{N_c-2}(N_c-k)t_k\pa_k
f_{tree,d}\over (\pa_1 f_{tree,d})^{N_c-1}} ,
\eeq{2a22}
\beq
B_{N_c}^{(0)}={-(N_c-1) f_{tree,d}+\sum_{k=1}^{N_c-2}(N_c-1-k)t_k\pa_k
f_{tree,d}\over (\pa_1 f_{tree,d})^{N_c}} .
\eeq{2a23}

Since eqs. (\ref{2a20}), (\ref{2a18}) imply $B_1^{(0)}=F_1(B_2^{(0)})=0$,
it follows that $B_2^{(0)}$ must be a constant in $(t_1,...,t_{N_c-2})$
(otherwise we would have $F_1(B_2)=0$ for any $B_2$ which implies the
unphysical result: $W_u=0$ in eq. (\ref{2a11})). Now, eq. (\ref{2a18}) implies
$B_k^{(0)}=F_k(B_2^{(0)})$ and, therefore, we conclude that $B_k^{(0)}$,
$k=2,...,N_c$ are constants.

There is only a finite number of branches of $W_{tree,d}$. Therefore, to
find which $f_{tree,d}$ has all $B_k^{(0)}=$ constant, we insert $\pa_k
f_{tree,d}$ from eq. (\ref{2a21}) in eqs. (\ref{2a22}), (\ref{2a23}) and,
after taking their combination, we obtain
\beq
t_1+\sum_{k=2}^{N_c}kt_k B_k^{(0)}(\pa_1 f_{tree,d})^{k-1}=0,
\eeq{2a24'}
\beq
f_{tree,d}=t_1 \pa_1 f_{tree,d}+\sum_{k=2}^{N_c} t_k B_k^{(0)} (\pa_1
f_{tree,d})^k ,
\eeq{2a24}
where $t_{N_c-1}=t_{N_c}=1$ by convention.
Equation (\ref{2a24'}) implies that
\beq
\z\equiv \pa_1 f_{tree,d}=-{t_1\over 2B_2^{(0)} t_2}+{\cal O}(t_2^{-2}) ,
\eeq{2a25}
and thus $W_{tree,d}$ is in the perturbative branch in $1/m_2$, namely,
in $1/t_2$ (recall eq. (\ref{2a1})), since (\ref{2a24}), (\ref{2a25})
give~\footnote{
Actually, there are $N_c-1$ solutions to eq. (\ref{2a24'}); one of them is
the perturbative solution. As explained in section 13, this is the
physical branch where the adjoint matter decouples in the infinite mass
limit ($m_2\to\infty$).
We also expect to have other physical branches, where the adjoint matter
decouples when $m_k\to\infty$ for $k>2$. Nevertheless, all $N_c-1$ solutions
to eq. (\ref{2a24'}) give rise to the same $W_u$ by integrating in.}
\beq
f_{tree,d}=-{t_1^2\over 4B_2^{(0)} t_2}+{\cal O}(t_2^{-2}) .
\eeq{2a26}

The perturbative branch of $W_{tree,d}$ is unique (see section 13). In
this branch, from eqs. (\ref{n1110}), (\ref{n1111}) one finds
\beq
\l X+\sum_{k=2}^{N_c} km_k{(1-N_c)^{k-1}-1\over (1-N_c)^{k-1}}\Big({\pa
W_{tree,d}\over \pa (\l X)}\Big)^{k-1}=0 .
\eeq{2a27}
Using (\ref{2a1}), (\ref{2a24'}), (\ref{2a24}), (\ref{2a25})
and (\ref{2a27}) we conclude that
\beq
\sum_{k=2}^{N_c} kt_k\Big(B_k^{(0)}+{1-(1-N_c)^{k-1}\over
(1-N_c)^{k-1}}\Big)\z^{k-1}=0 ,
\eeq{2a28}
for any $t$, which implies
\beq
B_k^{(0)}={(1-N_c)^{k-1}-1\over (1-N_c)^{k-1}} .
\eeq{2a29}

We now assume that $f_{\D}\neq 0$, and denote by
$a_n(t_1,...,t_{N_c-2}) t_0^n$ the
lowest non-zero order term of (\ref{2a6}); we will show that the
$m_2\to\infty$ limit implies $a_n=0$ and, therefore, $f_{\D}=0$.
Recall that $B_k^{(0)}$ are the zero order terms of the $t_0$ expansion of
$B_k(t)$. To next order in $t_0$ we obtain
\beq
B_1(t)=\Big({N_c\over N_c-1}{f_d\over
t_2\z^2}\Big)^{N_c}+{t_0^{n+1}\over t_1\z^{2N_c}}\Big[na_n-{2N_c\over
N_c-1}f_d{na_n+t_1\pa_1 a_n\over t_1\z}\Big]+... ,
\eeq{2a30}
\ber
B_2(t)=B_2^{(0)}&+&\Big({N_c\over N_c-1}{f_d\over t_2\z^2}\Big)\nonumber\\
&+&{t_0^n\over t_2\z^2}\Big[t_2\pa_2 a_n-2\Big(
{na_n+t_1\pa_1 a_n\over t_1\z}\Big)\Big({N_c\over N_c-1}f_d+t_2\pa_2
f_{tree,d}\Big)\Big] \nonumber\\ &+& ... ,
\eer{2a31}
\beq
B_k(t)=B_k^{(0)}+{t_0^n\over \z^k}\Big[\pa_k a_n-{kB_k^{(0)}\z^{k-1}\over
t_1}(na_n+t_1\pa_1 a_n)\Big]+... , \qquad k=3,...,N_c-2 ,
\eeq{2a32}
\ber
B_{N_c-1}(t)=B_{N_c-1}^{(0)}&+&{t_0^n\over \z^{N_c-1}}\Big[
-(N_c-1)B_{N_c-1}^{(0)}{\z^{N_c-2}\over t_1}(na_n+t_1\pa_1 a_n)\nonumber\\
&+&(N_c-nb_1)a_n-\sum_{k=1}^{N_c-2} (N_c-k)t_k\pa_k a_n\Big]+... ,
\eer{2a33}
\ber
B_{N_c}(t)=B_{N_c}^{(0)}&+&{t_0^n\over \z^{N_c}}\Big[
-N_c B_{N_c}^{(0)}{\z^{N_c-1}\over t_1}(na_n+t_1\pa_1 a_n)\nonumber\\
&-&(N_c-1-nb_1)a_n+\sum_{k=1}^{N_c-2} (N_c-1-k)t_k\pa_k a_n\Big]
\nonumber\\ &+&... ,
\eer{2a34}
where ``...'' mean higher orders in $t_0$,\footnote{
Recall that $f_d$, appearing in $B_1,B_2$, depends on $t_0$ (see eq.
(\ref{2a5})).} and $\z$ is given in (\ref{2a25}).

Using eqs. (\ref{2a5}), (\ref{2a18}) and
(\ref{2a30})-(\ref{2a34}), we find that the $a_n(t_1,...,t_{N_c-2})$ must
satisfy:
\beq
t_2\pa_2 a_n - na_n-2B_2^{(0)}{t_2\over t_1}\z (na_n+t_1 \pa_1 a_n)=
C_2 {t_2^{n+1}\over t_1^n} \z^{2N_c-2(N_c-1)(n+1)} ,
\eeq{2a35}
\beq
\pa_k a_n - kB_k^{(0)}{\z^{k-1}\over t_1}(na_n+t_1\pa_1 a_n)=C_k
\Big({t_2\over t_1}\Big)^n \z^{k-2n(N_c-1)} , \qquad k=3,...,N_c-2 ,
\eeq{2a36}
\ber
(N_c-nb_1)a_n&-&\sum_{k=1}^{N_c-2} (N_c-k)t_k \pa_k a_n -
(N_c-1)B_{N_c-1}^{(0)}(na_n+t_1 \pa_1 a_n){\z^{N_c-2}\over t_1}\nonumber\\
&=&C_{N_c-1} \Big({t_2\over t_1}\Big)^n \z^{(1-2n)(N_c-1)} ,
\eer{2a37}
\ber
(1-N_c+nb_1)a_n&+&\sum_{k=1}^{N_c-2} (N_c-1-k)t_k \pa_k a_n -
N_c B_{N_c}^{(0)}(na_n+t_1 \pa_1 a_n){\z^{N_c-1}\over t_1}\nonumber\\
&=&C_{N_c} \Big({t_2\over t_1}\Big)^n \z^{N_c-2n(N_c-1)} ,
\eer{2a38}
where the constants $C_k$, $k=2,...,N_c$ are defined as follows.
$C_2$ is proportional to the next to leading order coefficient in the
expansion of $B_1$ in powers of $B_2-B_2^{(0)}$ (recall eq. (\ref{2a18})):
\beq
B_1-(B_2-B_2^{(0)})^{N_c}\sim C_2(B_2-B_2^{(0)})^{(N_c-1)(n+1)}+... .
\eeq{2a39}
$C_k$, $k=3,...,N_c$ are proportional to the coefficients of the leading
order terms in the expansion of $B_k-B_k^{(0)}$ in powers of $B_2-B_2^{(0)}$
(recall eq. (\ref{2a18})):
\beq
B_k-B_k^{(0)}\sim C_k(B_2-B_2^{(0)})^{(N_c-1)n}+... , \qquad k=3,...,N_c .
\eeq{2a39'}
The solution of eqs. (\ref{2a35})-(\ref{2a38}) is
\beq
a_n=\Big({t_2\over t_1 \z^{2(N_c-1)}}\Big)^n \sum_{k=2}^{N_c} C_k \z^k t_k.
\eeq{2a39''}

It is now time to use the limit $m_2\to\infty$, namely, $t_2\to\infty$
(recall eq. (\ref{2a1})) to show that $a_n=0$.
In this limit $f_{\D}(t_2\to\infty)\to 0$ and,
therefore, $a_n(t_2\to\infty)\to 0$. Recall the  perturbative behavior
of $\z(1/t_2)$ in eq. (\ref{2a25}), we find that eq. (\ref{2a35}) implies
$C_2=0$, eq. (\ref{2a36}) implies $C_k=0$, $k=3,...,N_c-2$, eq.
(\ref{2a37}) implies $C_{N_c-1}=0$, and eq. (\ref{2a38}) implies
$C_{N_c}=0$; to summarize:
\beq
C_k=0, \qquad k=2,...,N_c .
\eeq{2a40}
Inserting (\ref{2a40}) in eq. (\ref{2a39''}) we find
\beq
a_n=0 .
\eeq{2a41}
Therefore, $f_{\D}=0$ (recall eq. (\ref{2a6})) and we conclude that, for
$N_c>3$, $W_{\D}=0$ on the $SU(N_c)$ vacua branch where $W_{tree,d}$ is the
one perturbative in $1/m_2$.

We now consider the $N_c=3$ case. For $SU(3)$, eq. (\ref{n113}) reads:
\beq
W_{tree,d}\equiv \Big[m_2\tr\P^2+m_3\tr\P^3
+\l Q\P\bQ\Big]_{\langle\P\rangle} .
\eeq{2a42}
We find $W_{tree,d}=\tau f_{tree,d}(t_1)$,  where $\tau$ and $t_1$ are
given by eqs. (\ref{2a2}) and (\ref{2a1}), respectively, with
$N_c=3$,~\footnote{Note that, unlike $N_c\neq 3$, here $\tau$ and
$t_0,t_1$ depend on $m_2$.} and in the the nonperturbative branch
\beq
f_{tree,d}(t_1)={8\over 9}+{2\over 3}t_1 ,
\eeq{2a43}
while in the perturbative branch
\beq
f_{tree,d}(t_1)={4\over 9}\Big[1-{3\over 2}t_1 \pm (1-t_1)^{3/2}\big] .
\eeq{2a44}
Indeed, in the perturbative branch~\footnote{
We consider both $\pm$ possibilities (the $N_c-1=2$ solutions of eq.
(\ref{2a24'}) with $N_c=3$) as the ``perturbative branch'' because
they give rise to the same $W_u$ (see the discussion for $N_c>3$); so we
may choose to work with the one obeying $W_{tree,d}(m_2\to\infty)\to 0$.},
$W_{tree,d}\to 0$ when $m_2\to\infty$.
We now follow eqs. (\ref{2a3})-({\ref{2a6}). Since $W_{\D}(\L\to 0)\to 0$,
and as we trust the instanton expansion in the Higgs phase, $f_{\D}$ is
holomorphic in $t_0$, defined in eq. (\ref{2a1}) with $N_c=3$, $b_1=5$, and
imposing also $W_{\D}(m_2\to\infty)\to 0$ we find
\beq
f_{\D}(t_0,t_1)=\sum_{n=1}^{\infty} a_n(t_1) t_0^n,
\eeq{2a45}
\beq
a_n(t_1)t_1^{(5n-3)/2}(t_1\to 0)\to 0 .
\eeq{2a45'}

Following eqs. (\ref{2a8})-(\ref{2a29}), with $N_c=3$, we find that imposing
the $\L\to 0$ behavior and a discrete set of vacua give rise to the
unphysical result $W_u=0$ in the nonperturbative branch (\ref{2a43}),
while for the perturbative branch ((\ref{2a44}) with the minus sign),
assuming $f_{\D}\neq 0$ and denoting by $a_n(t_1) t_0^n$ the lowest order
non-zero term in (\ref{2a45}), we obtain that
eqs. (\ref{2a30})-(\ref{2a34}) are being replaced with
\beq
B_1=\Big({3\over 2}{f_d\over \z^2}\Big)^3+t_0^{n+1}\Big({na_n\over t_1
\z^6} \Big)+... ,
\eeq{2a46}
\beq
B_2={3\over 2}+\Big({3\over 2}{f_d\over \z^2}\Big)+{t_0^n\over \z^2}\Big[
(3-5n)a_n-2t_1 a'_n-{3\z\over t_1}(na_n+t_1 a'_n)\Big]+... ,
\eeq{2a47}
\beq
B_3={3\over 4}+{t_0^n\over \z^3}\Big[
(5n-2)a_n+t_1 a'_n-{9\z^2\over 4t_1}(na_n+t_1 a'_n)\Big]+... .
\eeq{2a48}
We thus find that the expansion of $B_1$ and $B_3$ in terms of
$B_2-B_2^{(0)}=B_2-3/2$  is
\ber
B_1&=&\Big(B_2-{3\over 2}\Big)^3+\Big(B_2-{3\over 2}\Big)^{2(n+1)} 3^{n+1}
\z^{4n-2} t_1^n \Big[ (6n-3)a_n+2t_1 a'_n\nonumber\\
&+&{3\z\over t_1}(na_n+t_1 a'_n)\Big]+... ,
\eer{2a49}
\ber
B_2&=&{3\over 4}+\Big(B_2-{3\over 2}\Big)^{2n} 3^n \z^{4n-3} t_1^n \Big[
(5n-2)a_n+t_1 a'_n-{9\z^2\over 4t_1}(na_n+t_1 a'_n)\Big]\nonumber\\
&+&... .
\eer{2a50}
Now, the condition (\ref{2a18}) implies
\beq
(3-6n)a_n-2t_1 a'_n-{3\z\over t_1}(na_n+t_1 a'_n)=C_2 {\z^{2-4n}\over
t_1^n} ,
\eeq{2a51}
\beq
(5n-2)a_n+t_1 a'_n-{9\z^2\over 4t_1}(na_n+t_1 a'_n)=C_3{\z^{3-4n}\over
t_1^n} .
\eeq{2a52}
The solution of eqs. (\ref{2a51}), (\ref{2a52}) is
\beq
a_n={1\over (t_1\z^4)^n}(C_2\z^2+C_3\z^3) ,
\eeq{2a53}
where $\z$ is given in (\ref{2a25}) with $t_2=1$ .
So finally, we got in eq. (\ref{2a53}) the result (\ref{2a39''}) with
$N_c=3$ and $t_2=t_3=1$.

Finally, using the $m_2\to\infty$ limit which implies (\ref{2a45'}), we find
that $C_2=C_3=0$, and eq. (\ref{2a53}) implies $a_n=0$. Therefore,
$f_{\D}=0$, and also for $N_c=3$ on the $SU(3)$ vacua
branch, where $W_{tree,d}$ is the one perturbative in $1/m_2$, we conclude
that $W_{\D}=0$.

\end{document}